%% file: toloba_PAPERIII.tex
\documentclass[preprint2]{emulateapj}
\usepackage{multirow}

\newcommand{\Reff}{$R_e$}
\newcommand{\Rvirial}{$R_{\rm virial}$}
\newcommand{\ATLAS}{ATLAS$^{\rm 3D}$}
\newcommand{\kms}{km~s$^{-1}$}
\newcommand{\sigmae}{$\sigma_e$}
\newcommand{\Vrot}{$V_{\rm rot}$}
\newcommand{\vs}{$V_{\rm rot}/\sigma_e$}

\newcommand{\vsi}{$V/\sigma$} 
\newcommand{\vse}{$(V/\sigma)_e$} 
\newcommand{\lambdae}{$\lambda_{\rm Re}$}
\newcommand{\lambdaec}{$\lambda^*_{\rm Re}$}
\newcommand{\lambdaee}{$\lambda_{\rm Re/2}$}
\newcommand{\lambdaeec}{$\lambda^*_{\rm Re/2}$}
\newcommand{\lambdaeed}{$\lambda^{\rm 1D}_{\rm Re/2}$}
\newcommand{\lambdaed}{$\lambda^{\rm 1D}_{\rm Re}$}
\newcommand{\mue}{$\langle \mu_e \rangle$}

\newcommand{\reduceme}{\mbox{R\raisebox{-0.35ex}{E}D%
\hspace{-0.05em}\raisebox{0.85ex}{uc}\hspace{-0.90em}%
\raisebox{-.35ex}{{m}}\hspace{0.05em}E}}

\shorttitle{Structural and Kinematic Properties of Virgo cluster dEs. Rotation versus Pressure Support}
\shortauthors{Toloba et al.}

\begin{document}

\title{Stellar Kinematics and Structural Properties of Virgo Cluster Dwarf Early-Type Galaxies from the SMAKCED Project~III. Rotation versus Pressure Support.}


\author{E. Toloba\altaffilmark{1,2}\footnote{Fulbright Postdoctoral Fellow}}\email{toloba@ucolick.org}

\author{P.~Guhathakurta\altaffilmark{1}}

\author{A.~Boselli\altaffilmark{3}}

\author{R.~F.~Peletier\altaffilmark{4}}

\author{E.~Emsellem\altaffilmark{5,6}}

\author{T.~Lisker\altaffilmark{7}}

\author{G.~van~de~Ven\altaffilmark{8}}

\author{J.~D.~Simon\altaffilmark{2}}

\author{J.~Falc\'on-Barroso\altaffilmark{9,10}}

\author{J.~J.~Adams\altaffilmark{2}}

\author{A.~J.~Benson\altaffilmark{2}}

\author{S.~Boissier\altaffilmark{3}}

\author{M.~den Brok\altaffilmark{11}}

\author{J.~Gorgas\altaffilmark{12}}

\author{G.~Hensler\altaffilmark{13}}

\author{J.~Janz\altaffilmark{14}}

\author{E.~Laurikainen\altaffilmark{15,16}}

\author{S.~Paudel\altaffilmark{17}}

\author{A.~Ry\'s\altaffilmark{9,10}}

\author{H.~Salo\altaffilmark{15}}

\affil{$^1$UCO/Lick Observatory, University of California, Santa Cruz, 1156 High Street, Santa Cruz, CA 95064, USA}
\affil{$^2$Observatories of the Carnegie Institution for Science, 813 Santa Barbara Street, Pasadena, CA 91101, USA}
\affil{$^{3}$Aix Marseille Université, CNRS, LAM (Laboratoire d'Astrophysique de Marseille) UMR 7326, 13388, Marseille, France}
\affil{$^{4}$ Kapteyn Astronomical Institute, Postbus 800, 9700 AV Groningen, The Netherlands}
\affil{$^5$European Southern Observatory, Karl-Schwarzschild-Str. 2, 85748, Garching, Germany}
\affil{$^6$Universit\'e Lyon 1, Observatoire de Lyon, Centre de Recherche Astrophysique de Lyon and Ecole Normale Sup\'erieure de Lyon, 9 Avenue Charles Andr\'e, F-69230, Saint-Genis Laval, France}
\affil{$^7$ Astronomisches Rechen-Institut, Zentrum f${\rm \ddot{u}}$r Astronomie der Universit${\rm \ddot{a}}$t Heidelberg, M${\rm \ddot{o}}$nchhofstra${\rm \ss}$e 12-14, D-69120 Heidelberg, Germany}
\affil{$^8$ Max Planck Institute for Astronomy, K$\ddot{\rm o}$nigstuhl 17, 69117 Heidelberg, Germany}
\affil{$^9$ Instituto de Astrof\'{i}sica de Canarias, V\'{i}a L\'{a}ctea s$/$n, La Laguna, Tenerife, Spain}
\affil{$^{10}$ Departamento de Astrof\'{i}sica, Universidad de La Laguna, E-38205, La Laguna, Tenerife, Spain}
\affil{$^{11}$ Department of Physics and Astronomy, University of Utah, Salt Lake City, UT 84112, USA}
\affil{$^{12}$ Departamento de Astrof\'{i}sica y Ciencias de la Atm\'osfera, Universidad Complutense de Madrid, 28040, Madrid, Spain}
\affil{$^{13}$ University of Vienna, Department of Astrophysics, T${\rm \ddot{u}}$rkenschanzstra${\rm \ss}$e 17, 1180 Vienna, Austria}
\affil{$^{14}$ Centre for Astrophysics and Supercomputing, Swinburne University, Hawthorn, VIC 3122, Australia}
\affil{$^{15}$ Division of Astronomy, Department of Physics, P.O. Box 3000, FI-90014 University of Oulu, Finland}
\affil{$^{16}$ Finnish Center for Astronomy with ESO (FINCA), University of Turku, Finland}
\affil{$^{17}$ Korea Astronomy and Space Science Institute, Daejeon 305-348, Republic of Korea}

\begin{abstract}

We analyze the stellar kinematics of 39 dwarf early-type galaxies (dEs) in the Virgo cluster.
Based on the specific stellar angular momentum \lambdae\ and the ellipticity, we find 11 slow rotators and 28 fast rotators. The fast rotators in the outer parts of the Virgo cluster rotate significantly faster than fast rotators in the inner parts of the cluster. Moreover, 10 out of the 11 slow rotators are located in the inner $3^{\circ}$ ($D < 1$~Mpc) of the cluster. The fast rotators contain subtle disky structures that are visible in high-pass filtered optical images, while the slow rotators do not exhibit these structures. In addition, two of the dEs have kinematically decoupled cores and four more have emission partially filling in the Balmer absorption lines. 
These properties suggest that Virgo cluster dEs may have originated from late-type star-forming galaxies that were transformed by the environment after their infall into the cluster. The correlation between \lambdae\ and the clustercentric distance can be explained by a scenario where low luminosity star-forming galaxies fall into the cluster, their gas is rapidly removed by ram pressure stripping, although some of it can be retained in their core, their star-formation is quenched but their stellar kinematics are preserved. After a long time in the cluster and several passes through its center, the galaxies are heated up and transformed into slow rotating dEs. 

\end{abstract}


\section{Introduction}

Early-type galaxies (ETGs) are characterized by a smooth surface brightness distribution, small amounts of interstellar gas and dust, and red colors which indicate the presence of an old stellar population. The ETG galaxy class is composed of two morphological classes, ellipticals (Es) and lenticulars (S0s). This family of galaxies is often referred to as quenched or quiescent galaxies, in contrast to the late-type galaxy class that is often referred to as star-forming galaxies.

Dwarf early-type galaxies are the low luminosity ($M_B \gtrsim -18$) and low surface brightness ($\mu_B \gtrsim 22$~mag~arcsec$^{-2}$) population of the ETG class. The term dE has traditionally been used to refer to dwarf elliptical galaxies. In this work we loosely use dE for all quiescent dwarf galaxies ($M_B \gtrsim -18$, $\mu_B \gtrsim 22$~mag~arcsec$^{-2}$), including dwarf elliptical and dwarf lenticular (dS0) galaxies.

The dE galaxy class spans a wide range of internal properties. Despite their apparent simplicity, several dEs have complex structures beneath their smooth light distribution. These structures are in form of spiral arms, disks, and/or irregular features \citep[e.g.,][]{Jerjen00,Barazza02,Geha03,Graham03,DR03,Lisk06a,Ferrarese06,Janz12,Janz14}. The kinematics and stellar populations of dEs are also complex. Dwarf early-type galaxies with similar photometric properties can have different rotation speeds \citep[][]{Ped02,SimPrugVI,Geha02,Geha03,VZ04,Chil09,etj09,etj11,Rys13,Rys14}, and also different stellar populations \citep[e.g.][]{Mich08,Paudel2010,Koleva11}

Dwarf early-type galaxies are found in high density environments and are rarely seen in isolation \citep[][]{Gavazzi10,Geha12}. Moreover, dEs dominate in number the population of high density environments \citep{Dress80,Sand85,Bing88}.

This morphological segregation and the complexity of the internal properties suggest that the progenitors of dEs are late-type star-forming galaxies that were transformed by the environment. The two most widely discussed mechanisms indicate that this transformation occurs through gravitational tidal heating \citep[i.e., harassment, ][]{Moore98,Mast05} or through hydrodynamical interactions with the intracluster medium \citep[i.e., ram-pressure stripping][]{LinFab83,Boselli08a,Boselli08b}.

With the goal of understanding the physical processes that form dEs, we have begun the SMAKCED\footnote{http://smakced.net} (Stellar content, MAss and Kinematics of Cluster Early-type Dwarf galaxies) project, a new spectroscopic and photometric survey of dEs in the Virgo cluster, the nearest dense galaxy cluster. This paper is part of a series in which we analyze the structural and kinematic properties of dEs in the Virgo cluster. Details of the photometric and spectroscopic data sets are described in \citet[][, Paper~I in this series]{Janz14,etj14a} and Toloba et al. (2014, in press --- hereafter referred to as Paper~II). This paper, the third in the series, is focused on the study of stellar kinematics trends as a function of projected distance to the center of the Virgo cluster. 

In a gradual transformation of the internal properties of galaxies from environmental effects, trends between these internal properties and the density of the environment are expected. Galaxy clusters are the ideal laboratories to study these trends because the density gradient between the inner and outer parts of the cluster is usually large. However, these trends can be investigated only in galaxy clusters that are still under assembly, as is the case for Virgo \citep{BG06,Gavazzi13,Boselli14}, otherwise the trends disappear as galaxies evolve. Morphologically, nucleated and rounder dEs appear concentrated in the high density regions of the Virgo cluster while non-nucleated dEs are found everywhere in the cluster \citep{Lisk07,Lisk09}. Kinematically, \citet{etj09} found a hint of a correlation between the rotation support of the dEs and their location within the Virgo cluster but with a marginal statistical significance. To give these interesting results a stronger statistical basis, we have more than doubled the sample of \citet{etj09}, 2.2~times larger, we have done a more robust measurement of the rotation speed modeling the rotation curves of the galaxies, we have improved the $V/\sigma$ measurements using integrated values within the \Reff, and we have combined that kinematic information with structural properties.

This paper is organized as follows. In Section \ref{data}, we describe the sample of galaxies used in this analysis, the instrumental setups used in the spectroscopic observations, and the main steps in the reduction. In Section \ref{kin_measurements}, we present the measurement of the stellar kinematics. In Section \ref{vs_section}, we analyze the nature of the internal kinematics as a function of projected distance from the center of the Virgo cluster. In Section \ref{Sm_comp} we compare the structural properties and stellar kinematics of the SMAKCED dEs with those of dwarf star-forming galaxies. In Section \ref{discussion} we discuss the possible formation scenarios for dEs in the Virgo cluster. Finally, in Section \ref{concl} we summarize our findings and conclusions.

\section{The Data}\label{data}

This paper uses optical spectroscopy collected as part of the SMAKCED project. We present here a brief summary of the sample selection, observations, and data reduction. For more details see \citet{etj11} and Paper~II.

\subsection{Sample}\label{sample}

This paper is focused on the long-slit spectroscopic observations of 39 dEs in the Virgo cluster. The sample is representative of the full population of Virgo cluster dEs in the magnitude range $-19.0 < M_r < -16.0$ (see Paper~II).

These 39 dEs are selected prioritizing high surface brightness galaxies from the subsample of 121 dEs in the Virgo cluster analyzed in the $H$ band by \citet{Janz12,Janz14}. The galaxies are selected from the Virgo Cluster Catalog \citep[VCC,][]{Bing85} with updated memberships based on radial velocities \citep{Lisk06a}.  Their selection is based on their early-type morphology and low luminosity; see details in \citet{Janz14} and Papers~I and~II. 

Eighteen of the dEs in the sample were observed as part of the MAGPOP-ITP collaboration (Multiwavelength Analysis of Galaxy POPulations-International Time Program) and already presented in \citet{etj09,etj11,etj12}. In this series of papers we reanalyze the stellar kinematics of these 18 galaxies, changing the radial binning scheme from a minimum of 1~pixel to 3~pixels (see Section \ref{kin_measurements}). The remaining 21 dEs are new observations presented here for the first time.

\begin{figure}
\centering
\includegraphics[angle=-90,width=8.5cm]{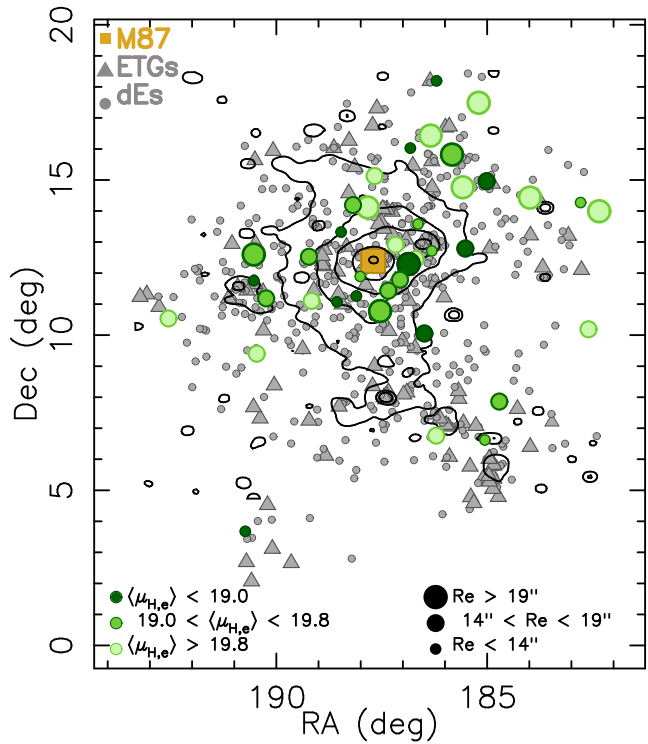}
\caption{Spatial distribution of the SMAKCED dEs within the Virgo cluster. Triangles indicate massive ETGs, and dots indicate dEs. The yellow square indicates the position of M87, considered to be the center of the Virgo cluster. The green dots highlight the 39 SMAKCED dEs analyzed here. The sizes of the green dots are related to the $H$ band \Reff~ of the dEs. The intensity of the green color is related to the $H$ band \mue~ of the dE (measured in mag~arcsec$^{-2}$). The structural parameters of the SMAKCED dEs can be found in Paper~II. The black contours indicate the X-ray diffuse emission of the cluster from \citet{Bohringer94}.}
\label{RADec}
\end{figure}

The spatial distribution of the 39 SMAKCED dEs within the Virgo cluster is shown in Figure \ref{RADec}. These 39 dEs cover a wide range of projected distances from M87. The Virgo cluster has a complex structure of subgroups \citep{Bing93}. The larger subgroup, Virgo A, is centered on M87 (${\rm Dec} \gtrsim 9^{\circ}$) and the smaller subgroup Virgo B is centered on M49 (${\rm Dec} < 9^{\circ}$). The majority of the SMAKCED dEs are part of Virgo A, only 4 of them are part of Virgo B \citep{Gavazzi99}. 
The sizes and intensities of the green dots in Figure \ref{RADec} indicate the half-light radius (\Reff) and mean surface brightness within the \Reff\ (\mue) of the galaxies. The \Reff\ and surface brightness of the SMAKCED dEs do not depend on their position within the cluster, in the same way that the full population of dEs do not depend on that either \citep[e.g.][]{Lisk07,Lisk09}.

\subsection{Observations and data reduction}\label{obs}

The observations were conducted at El Roque de los Muchachos Observatory (Spain) and the European Southern Observatory (ESO; Chile). 
Ten out of the 39 dEs were observed at the INT 2.5~m telescope, 26 dEs were observed at the WHT 4.2~m telescope, and the remaining 3 dEs were observed at the VLT 8~m telescope.
The exposure times varied from 1 to 3 hours depending on the brightness of the dE.

The observations at the INT were carried out using the IDS spectrograph with the 1200~l~mm$^{-1}$ grating covering the wavelength range 4600$-$5600~\AA. The spectral resolution obtained, using a slit width of $2''$, is 1.6~\AA~(FWHM).

The observations at the WHT were carried out using the double-arm spectrograph ISIS, which allowed us to observe two spectral ranges simultaneously. The blue setup used the 1200~l~mm$^{-1}$ grating and covered the wavelength range 4200$-$5000~\AA. The red setup used the 600~l~mm$^{-1}$ grating and covered the wavelength range 5500$-$6700~\AA. The spectral resolution obtained, using a slit width of $2''$, is 1.4 and 3.2~\AA~(FWHM) in the blue and red setup, respectively.

The observations at the VLT were carried out using the FORS2 spectrograph with the 1400V grism and covered the wavelength range 4500$-$5600~\AA. The spectral resolution obtained, using a slit width of $1.3''$, is 2.7~\AA~(FWHM).

The data were reduced following the standard procedure for long-slit spectroscopy using the package \reduceme~ \citep{Car99}. Here we summarize  the main steps; for more details see \citet{etj11} and Paper~II. The main steps consisted of bias and dark current subtraction, flat fielding, and cosmic ray cleaning. The spectra were then spatially aligned, wavelength calibrated, sky subtracted, and flux calibrated using the response function derived from our observed flux standards.

\input{kinematics_table.tex}

\section{Kinematic Measurements}\label{kin_measurements}

The line-of-sight radial velocity ($V$) and velocity dispersion ($\sigma$) are estimated along the galaxy's major axis using the penalized pixel-fitting software pPXF \citep{PPXF}. This software fits the galaxy spectrum with a model created from a linear combination of stellar templates that best reproduce the galaxy spectrum, allowing different weights for each template. The stellar templates used in this work are stars selected from the MILES stellar library \citep{SB06lib,MILEScen,FB11MILES} that have been observed with the same instrumental setup as the galaxies. The stars were observed with the telescope defocused to make them homogeneously fill the slit as the galaxies do. To be able to reproduce the stellar populations of the galaxies, the stellar templates cover a variety of spectral types and luminosity classes (B9, A0, A5V, G2III, G2V, G8III, G9III, K0I, K1V, K2III, K3III, K4III, M2III).

The three dEs observed at the VLT --- VCC 940, VCC 1684, and VCC 2083 --- do not have stellar templates observed under the same conditions as the galaxies, so we used stars from the ELODIE stellar library \citep{Prug07} as templates instead. These spectra, which have a resolution of R $\sim 10000$, are convolved with a Gaussian function whose width is equal to the quadratic difference between the instrumental resolution of the galaxies observed at the VLT and the ELODIE resolution. Details about the potential differences in the kinematic estimations due to the different techniques used are described in Paper~II.

The reduced two dimensional spectrum is spatially co-added following the size and S/N thresholds described in \citet{etj11} and in Paper~II. The minimum size over which to co-add spectra is 3 pixels, which is comparable to the average seeing affecting the observations. The minimum S/N threshold is chosen based on the simulations described in \citet{etj11}. These simulations use a large range of stellar populations, different radial velocities consistent with being Virgo cluster members, and velocity dispersions in the range of dwarf galaxies ($20-60$~\kms). These simulations indicate that the reliability of the measured radial velocity is not guaranteed for spectra with S/N below 10~\AA$^{-1}$, and the same happens for velocity dispersion estimations with S/N~$<15$~\AA$^{-1}$. Thus, we require S/N~$\geq 10$~\AA$^{-1}$ to measure $V$ and S/N~$\geq 15$~\AA$^{-1}$ to measure $\sigma$.
To derive the stellar kinematics for those galaxies with several masked regions, as a result of strong sky lines or emission lines, the minimum S/N required is $\geq 15$~\AA$^{-1}$ for $V$ and $\geq 25$~\AA$^{-1}$ for $\sigma$.

The parameter \vsi~ is used to quantify the dynamical support of the galaxies and the anisotropy of their velocity field \citep[e.g.][]{Binn05}. This parameter, based on the virial theorem, takes integrated quantities not easily estimated observationally. However, \citet{Em07} and \citet{Cappellari07} show that some corrections can be made to get the observational values as close as possible to the virial theorem.

Observationally, we obtain \vs, where \Vrot~ is the velocity at the \Reff~ measured in the best fit {\it Polyex} function to the rotation curve (Paper~II). 
The velocity dispersion \sigmae~ is measured by co-adding the spectra within the \Reff, so this measurement contains also the luminosity-weighted rotation.

To transform the measured long-slit \vs~ into the integrated \vsi, we simulate the two-dimensional distribution of the flux, $V$, and $\sigma$ based on the long-slit spectroscopic measurements. We use the ellipticity gradients from the $H$ band images presented in Paper~II. We smoothly decrease the rotation following a cosine function and make it zero along the minor axis, and we assume that the velocity dispersion profile is flat (see Paper~II), thus the values at each radius can be extrapolated in elliptical rings. These simulations are generated only in the regions where we have spectroscopic data. Then, we measure \vse~ in the simulation, which is the integrated two-dimensional \vsi~ measured within the \Reff. We find the best fit  between \vse~ and \vs

\begin{equation}\label{scale_vs_eqn}
(V/\sigma)_e = (0.39 \pm 0.02) V_{\rm rot}/\sigma_e
\end{equation}

\noindent the slope of this relation is smaller than the value found by \citet{Cappellari07} because they were comparing \vse\ to the maximum rotation speed instead of \Vrot, and the central velocity dispersion instead of \sigmae.

The main inconvenience of using \vsi~is that, as it is a luminosity-weighted parameter, it gives the highest weight to the central kinematics, which is not necessarily representative of the whole galaxy, as it is the case of  kinematically decoupled cores. To overcome this, \citet{Em07} introduced the $\lambda_{\rm R}$ parameter, which is an estimate of the specific stellar angular momentum of a galaxy, and it is defined as:

\begin{equation}\label{lamb_eqn}
\lambda_{\rm R} = \frac{\sum_{i=1}^N F_i R_i |V_i|}{\sum_{i=1}^N F_i R_i \sqrt{V_i^2+\sigma_i^2}}
\end{equation}

\noindent where $N$ is the number of spatial bins in the kinematic profiles, and $F_i$, $R_i$, $V_i$, and $\sigma_i$ are the flux, radius, rotation velocity, and velocity dispersion of the $i$th bin. The radii for the individual $V$ and $\sigma$ measurements are usually not the same in our data because of the two different S/N thresholds required to make these measurements. The parameter $R_i$ refers to the radius at which $V_i$ is measured. If there is a $\sigma$ measurement at that same radius, $\sigma_i$ is that value, if not, $\sigma_i$ is the average of the two closest values of $\sigma$ to the radius $R_i$. Some of the velocity dispersion profiles do not go as far out in radius as the rotation curves (this happens by definition because of the higher S/N required to measure $\sigma_i$; see Paper~II). In the region where $\sigma_i$ is not available, we use the uncertainty-weighted average of the velocity dispersion profile since the $\sigma$ profiles are generally flat \citep[see Paper~II and][for more flat velocity dispersion profiles of Virgo cluster dEs]{Geha02,Geha03,Chil09,Rys13}. For all the SMAKCED dEs, $\lambda_{\rm R}$ can be measured within half the \Reff~ (\lambdaeed), while only 25 of them can be measured within the \Reff~ (\lambdaed).
The uncertainties in \lambdaeed~ and \lambdaed~ are estimated by running 100 Monte Carlo simulations perturbing $V_i$ and $\sigma_i$ within a Gaussian function with a width equal to their uncertainties.

To transform the measured long-slit $\lambda^{\rm 1D}_{\rm R}$ into the integrated $\lambda_{\rm R}$ we use the same simulations produced for \vse. We find the best fit between $\lambda_{\rm R}$ and $\lambda^{\rm 1D}_{\rm R}$

\begin{equation}\label{scale_lamb_eqn}
\lambda_{\rm R} = (0.64 \pm 0.08) \lambda ^{\rm 1D}_{\rm R}
\end{equation}

\begin{figure}
\centering
\includegraphics[angle=-90,width=8.5cm]{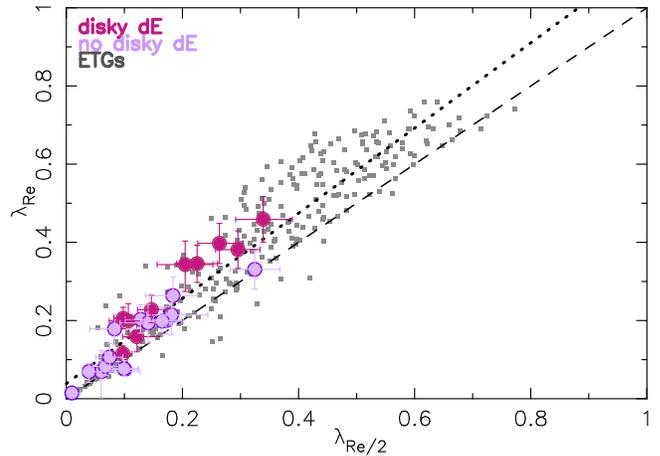}
\caption{Comparison between the stellar angular momentum measured within the \Reff~ ($\lambda_{\rm {Re}}$) and within the \Reff$/2$ ($\lambda_{\rm {Re/2}}$). Large dots are the SMAKCED sample of dEs, small squares are the \ATLAS sample of ETGs. Purple and red colors refer to the absence or presence of underlying disky structures observed in unsharp masked SDSS images by \citet{Lisk06a}. The dashed line indicates the $\lambda_{\rm {Re}} = \lambda_{\rm {Re/2}}$ relation, and the dotted line is the best linear fit to the data.}
\label{lambRevsRe2}
\end{figure}

Figure \ref{lambRevsRe2} shows a comparison between \lambdae~ and \lambdaee, obtained using Equation \ref{scale_lamb_eqn},  for the 25 dEs for which the radial extent reaches the \Reff. It also shows these values for the ETGs of the \ATLAS~ sample. The median ratio $\lambda_{\rm Re}/\lambda_{\rm Re/2}$ for the SMAKCED dEs is $1.38^{+0.31}_{-0.22}$, which is consistent with the ratio for the \ATLAS~ sample ($1.16^{+0.31}_{-0.16}$). It is expected that \lambdae~ is larger than \lambdaee~ because the rotation curves continue increasing beyond the \Reff~ \citep[e.g.][]{Beasley09,Geha10}. 
We fit a line to \lambdae~ vs. \lambdaee~ and use it to estimate the \lambdae~ value for those galaxies where the kinematic radial coverage does not reach the \Reff. The resulting \lambdae~ values are in good agreement with those measured by \citep{Rys14}.

\begin{figure}
\centering
\includegraphics[angle=-90,width=8.5cm]{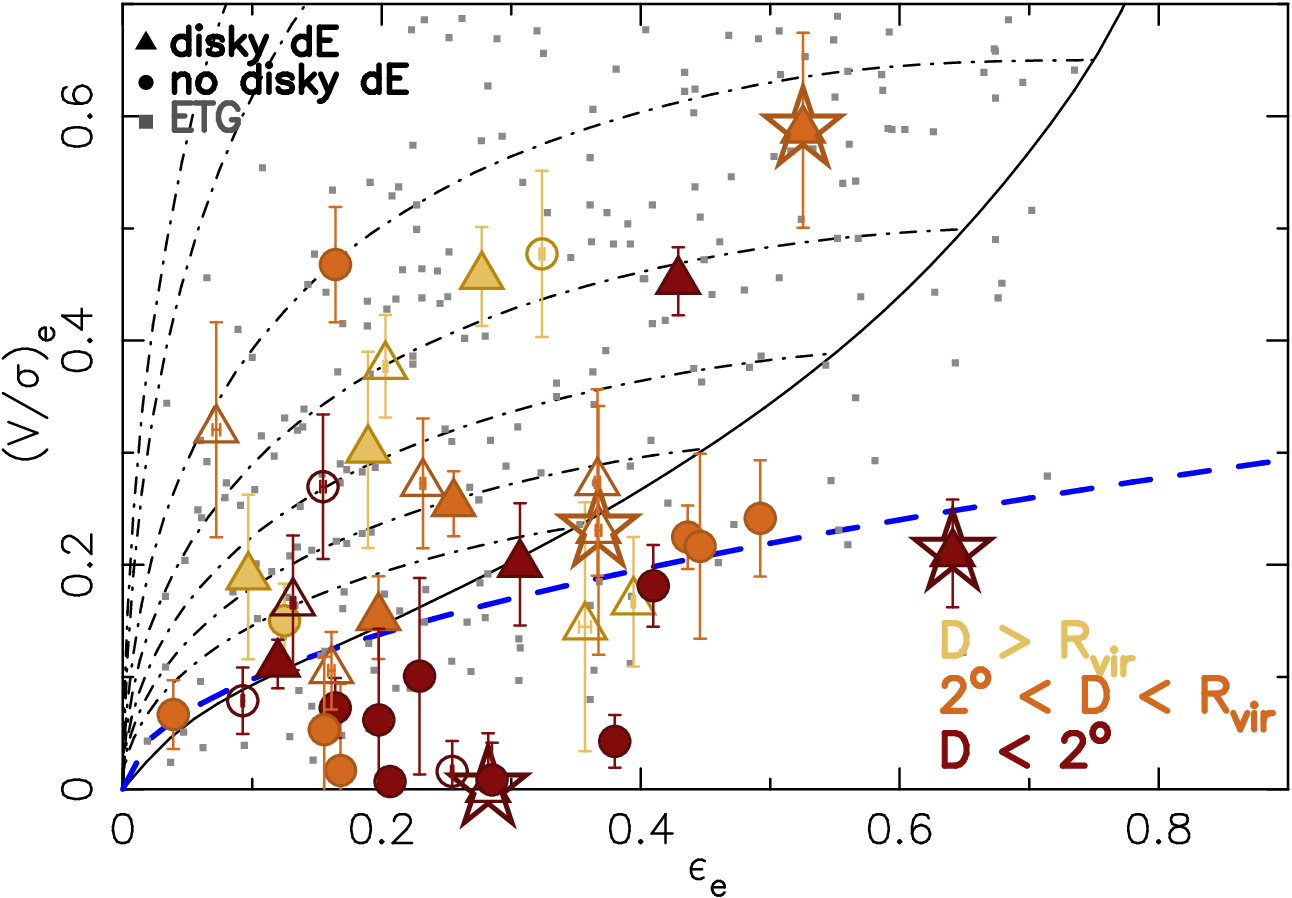}
\includegraphics[angle=-90,width=8.5cm]{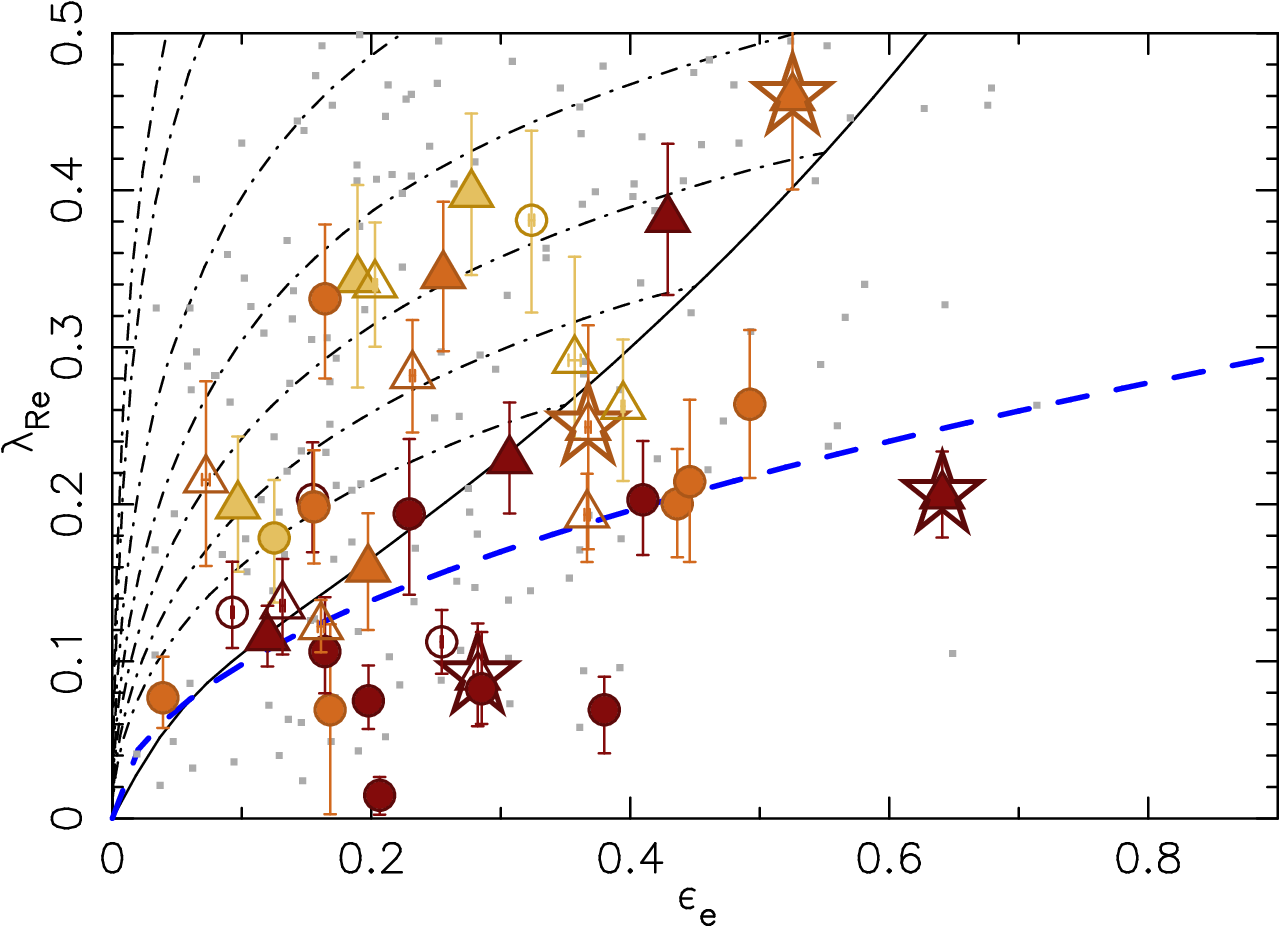}
\includegraphics[angle=-90,width=8.5cm]{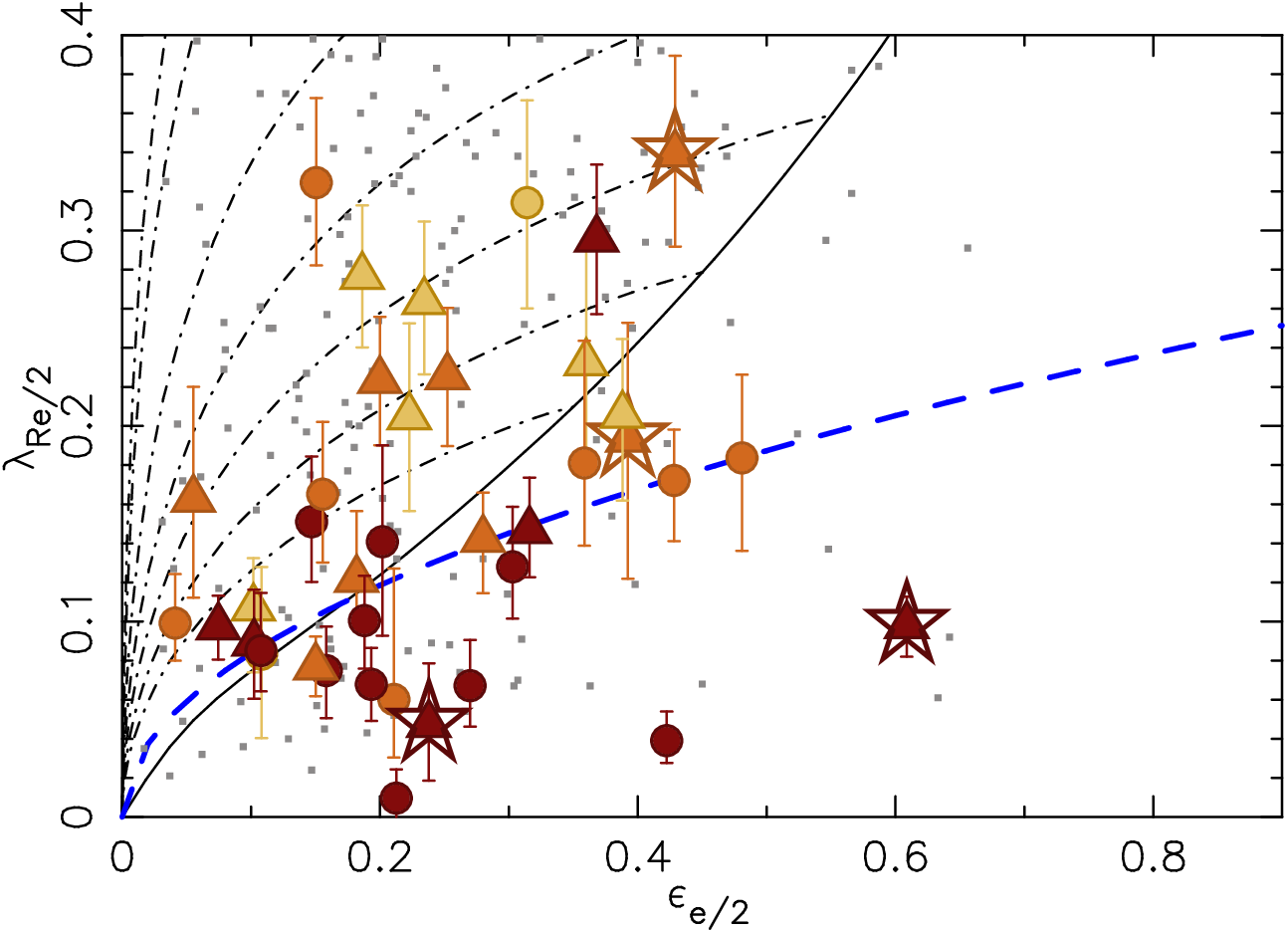}
\caption{Stellar angular momentum \vse, \lambdae\ and \lambdaee\ as a function of ellipticity. The red, orange, and yellow symbols are the SMAKCED dEs. The colors refer to their projected distance to the center of the cluster. The shape indicate the presence or absence of subtle disky structures seen in high-pass filtered SDSS images by \citet{Lisk06a}. The large open stars indicate the dEs for which we found some emission filling in the Balmer lines. The open circles and triangles indicate the dEs for which the kinematic radial coverage is shorter than the \Reff. 
The grey squares are the \ATLAS~ sample of ETGs \citep{Em11}. 
The black lines (solid and dot-dashed) are the models for anisotropic galaxies calculated with independent dynamical models by \citet{Em11}.
The black solid line is the expected relation for an edge-on ellipsoidal galaxy with an anisotropy of $\beta=0.65 \times \epsilon$ within an aperture of 1\Reff. The black dot-dashed lines indicate the effect of inclination on this relation. From top to bottom, the dot-dashed lines show the relation for galaxies with intrinsic ellipticities of 0.95, 0.85, 0.75, 0.65, 0.55, 0.45, and 0.35, and from right to left the inclination changes from edge-on (on the black solid line relation) to face-on (on the origin).
The blue dashed line indicates the separation between slow (below the line) and fast (above the line) rotators defined by \citep{Em07,Em11} as indicated in Equations \ref{eqn_vs_e},  \ref{eqn_lam_e}, and \ref{eqn_lam_ee}.}
\label{vs_e}
\end{figure}

\begin{figure*}
\centering
\includegraphics[angle=-90,width=11cm]{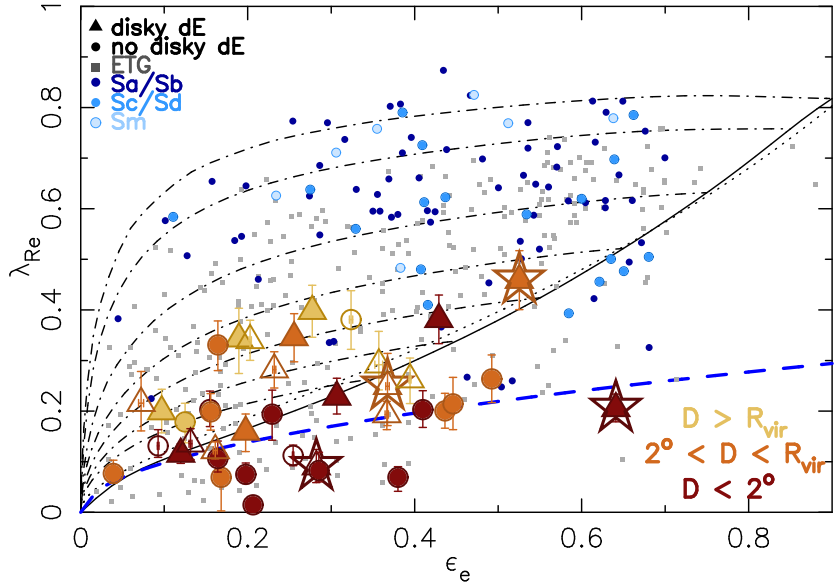}
\caption{Stellar angular momentum \lambdae~ as a function of ellipticity. The symbols, colors, and lines are as in Figure \ref{vs_e}. The dark and intermediate blue dots are the late-type galaxies from Falc\'on-Barroso et al. (in prep.). The light blue dots are the dwarf star-forming (Sm) galaxies from \citet{Adams14}. Star-forming galaxies and ETGs reach higher values of \lambdae\ than dEs. The parameter \lambdae\ seems to be related to the structural properties of the dEs and their position within the Virgo cluster.}
\label{lambdaRe_e}
\end{figure*}

\section{Rotation versus Pressure Support: Nature of Internal Kinematics}\label{vs_section}

The kinematic properties of ETGs depend on the local galaxy density, i.e. the number and spatial distribution of surrounding galaxies, which is expected for massive galaxies where mergers play an important role in their evolution \citep{Cappellari11}. However, the merging rate found for simulated low mass galaxies is extremely low \citep[e.g.,][]{deLucia06,Fakhouri10}. 

Low mass galaxies are expected to be affected by the hot intracluster medium (e.g; ram pressure stripping) and the possible multiple encounters with more massive galaxies (e.g.; harassment), neither of these effects are necessarily described by the number of galaxies that are surrounding the dEs today. Moreover, \citet{Boselli14} find no evidence that the properties of Virgo cluster dEs depend on the local galaxy density.

The internal kinematics of galaxies can be dominated by rotation (rotation support: the stars move in a preferential direction, as in disks of spiral galaxies) or by dispersion (pressure support: the stars move randomly with no preferential direction, as in the most massive ETGs). In this Section we explore the nature of the internal kinematics of dEs by analyzing the parameters \vse, \lambdae, and \lambdaee\ and whether these correlate with the projected distance from the center of the Virgo cluster. This projected distance is a proxy of the impact that the hot intracluster medium may have had on the galaxies. We use the projected distance to M87, which hosts the brightest $X$-ray source of the Virgo cluster \citep{Schindler99}, as a tracer of the hot gas density. 
The adopted distance for galaxies in the Virgo cluster is 16.5~Mpc \citep{Mei07}, and its virial radius, assuming $R_{virial}=R_{200}$, is 1.55~Mpc \citep{Ferrarese12}.

Figures \ref{vs_e} and \ref{lambdaRe_e} show the distribution of dEs in the \vse$- \epsilon_e$, \lambdae$-\epsilon_{e/2}$, and \lambdaee$-\epsilon_e$ planes. These planes, which relate the internal kinematics and the apparent flattening of the galaxies, are used to derived the anisotropy of galaxies and also to distinguish between fast rotators and slow rotators.
The definition of fast and slow rotators (FR and SR, respectively) introduced by \citet{Em07,Em11} is indicated as blue dashed lines and follows the relations:

\begin{equation}\label{eqn_vs_e}
(V/\sigma)_{e} = 0.310 \times \sqrt{\epsilon_e}
\end{equation}

\begin{equation}\label{eqn_lam_e}
\lambda_{\rm Re} = 0.310 \times \sqrt{\epsilon_e}
\end{equation}

\begin{equation}\label{eqn_lam_ee}
\lambda_{\rm Re/2} = 0.265 \times \sqrt{\epsilon_{e/2}}
\end{equation}

\noindent where $\epsilon_e$ and $\epsilon_{e/2}$ are the average ellipticities within the \Reff~ and \Reff$/2$, respectively.

The dEs span a large range in ellipticity, similar to the ellipticity range covered by more massive ETGs, but their $(V/\sigma)_{e}$, \lambdae, or \lambdaee\ do not reach the highest values reached by ETGs (note that Figure \ref{vs_e} is a zoom-in of Figure \ref{lambdaRe_e}). There are round and flattened slow rotating dEs (SR, below the blue dashed line), as well as round and flattened fast rotating dEs (FR, above the blue dashed line). Figure \ref{lambdaRe_e} also shows the distribution of late-type galaxies in this space of parameters (values from the integral field spectroscopy presented in Falc\'on-Barroso et al., in prep. for Sa, Sb, Sc, and Sd galaxies, and in \citet{Adams14}, for Sm galaxies). The absence of round ($\epsilon_e < 0.2$) dwarf star-forming galaxies (Sm) is a selection effect (face-on star-forming galaxies are preferentially avoided). Galaxies with Sa, Sb, Sc, and Sd morphology overlap with the ETGs with \lambdae$\gtrsim 0.4$ and reach even higher values than the ETGs. Dwarf star-forming galaxies, Sm, tend to show the highest values of \lambdae\ (\lambdae$\gtrsim 0.6$) reaching also values higher than those found in ETGs. This may indicate that low surface brightness star-forming galaxies tend to have higher values of \lambdae\ than high surface brightness star-forming galaxies, as expected from simulations \citep[e.g.][]{Boissier00}. This possible relation needs to be further investigated using complete samples of star-forming galaxies.

VCC~1684 is a highly flattened SR dE with an ellipticity larger than 0.6. Two ETGs in the \ATLAS~ sample, NGC~4473 and NGC~4550, have similar properties and both are classified as $2\sigma$ galaxies, i.e. galaxies with two two peaks in the velocity dispersion profile possibly caused by two counter rotating disks \citep{Kraj11}. They are both on the low mass side of the \ATLAS~ sample \citep[$\log {\rm M}_e \sim 10.2$~M$_{\odot}$;][]{Em11}, and have emission lines \citep{Sarzi06}. We do not have evidence for two velocity dispersion peaks in VCC~1684. If these features are not along the photometric major axis of the galaxy, the long-slit spectroscopy presented in this series of papers would have missed them. However, we do classify VCC~1684 as a poor fit (i.e. significant departures of the rotation curve from the best fit {\it Polyex} function) and asymmetric rotation curve (i.e. the approaching and receding sides of the rotation curve have different shapes; see Paper~II for details). In addition, we also find emission lines in VCC~1684. Its dynamical mass within the \Reff\ is within the average mass of the SMAKCED sample ($\log {\rm M}_e = 8.9$~M$_{\odot}$). These properties suggest that VCC~1684 is a good candidate to be a $2\sigma$ galaxy. Integral field spectroscopy of this dE will show whether the velocity dispersion map shows two peaks.

The segregation of triangles and dots in Figure \ref{vs_e} suggests that the internal kinematics of dEs are related to the presence or absence of residual disky structures observed beneath their smooth light distribution in optical images. The dEs with disky structures have, on average, \vse$=0.27\pm0.03$, \lambdaee$=0.18\pm0.02$, and \lambdae$=0.25\pm0.02$. The dEs without disky structures have, on average, \vse$=0.14\pm0.03$, \lambdaee$=0.13\pm0.02$, and \lambdae$=0.16\pm0.02$. This suggests that dEs with underlying disky structures are rotating faster than the dEs without underlying disky structures. While inclination affects the observed rotation speed of the galaxies, the disky structures are independent from it, i.e disks are better seen in edge-on galaxies but spiral arms and bars are better seen in face-on galaxies \citep{Lisk06a}.

The colors in Figures \ref{vs_e} and \ref{lambdaRe_e} indicate the projected clustercentric distance ($D$, assuming that M87 is in the center of the Virgo cluster). Dwarf early-types with darker colors tend to have lower values of \vse, \lambdae, and \lambdaee, i.e. the dEs in the inner regions of the cluster rotate slower than the dEs in the outer regions. Even though the sample of dEs is not complete, the selection function is not different for dEs in the inner and outer parts of the cluster.

\begin{figure}
\centering
\includegraphics[angle=-90,width=8.5cm]{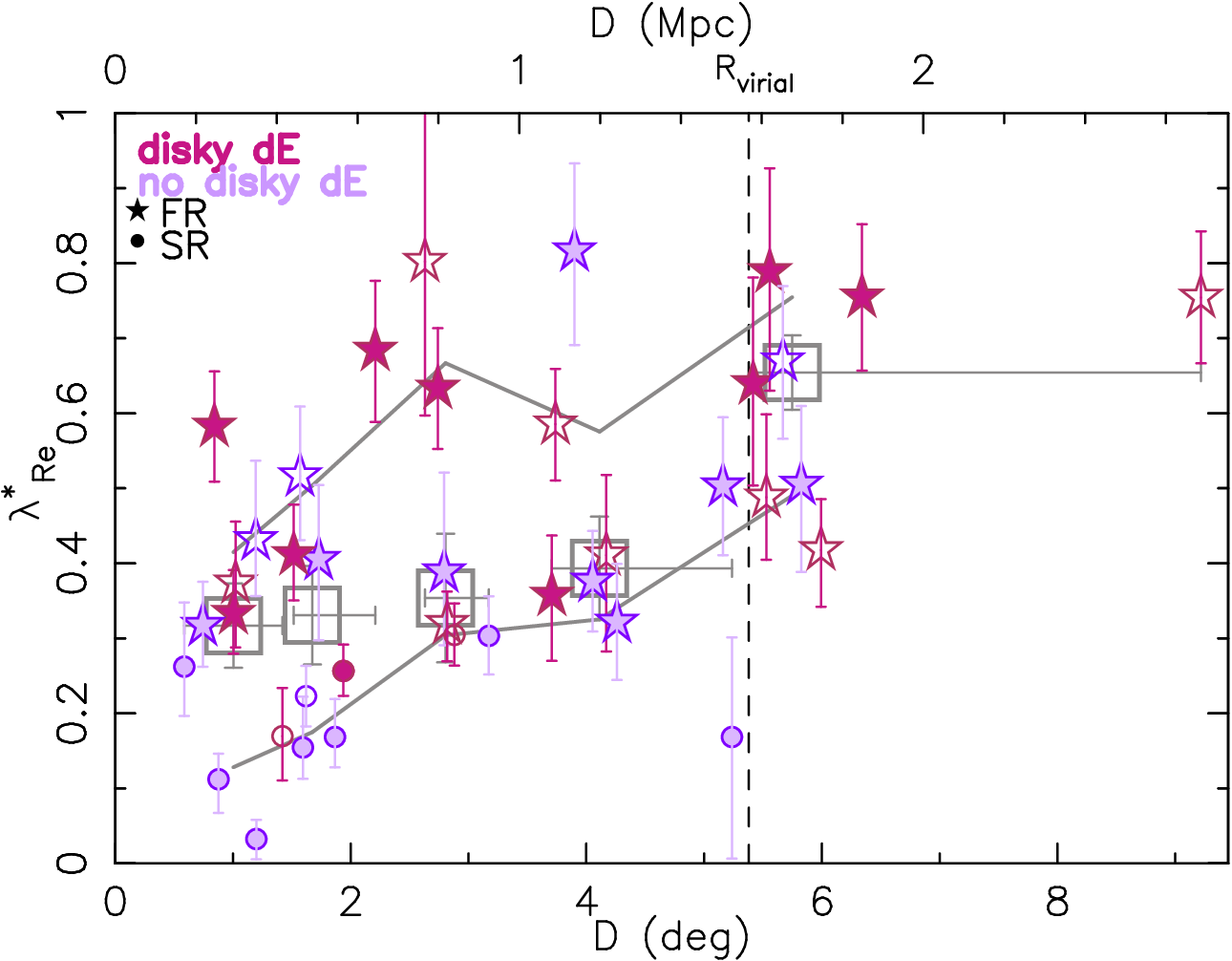}
\includegraphics[angle=-90,width=8.5cm]{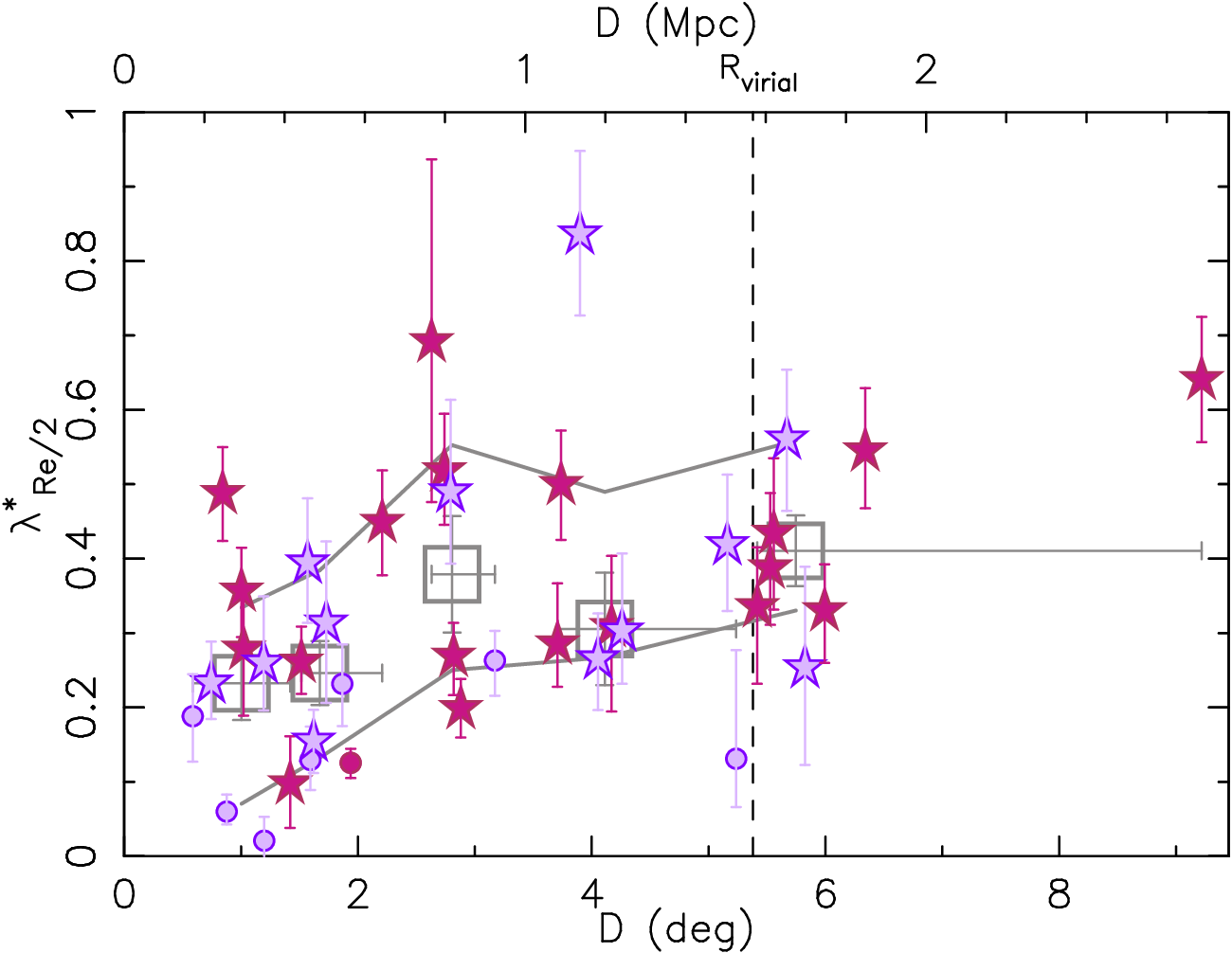}
\caption{Spatial distribution of SR and FR dEs within the Virgo cluster. The upper and lower panel show the specific stellar angular momentum normalized by the square root of the ellipticity following Equations \ref{eqn_lambdaec} and \ref{eqn_lambdaeec} as a function of the projected clustercentric distance. Colors are as in Figure \ref{lambRevsRe2}. The shape of the symbols indicate whether the galaxies are fast rotators (asterisks; they are at all distances from the center of Virgo) or slow rotators (dots; they are preferentially in the inner region of Virgo) based on the blue line shown in Figure \ref{lambdaRe_e}. The dashed vertical line indicates the virial radius of the cluster. The open and filled red and purple symbols are as in Figure \ref{lambdaRe_e}. The grey open squares and solid lines are the median and $16^{\rm th}$ and $84^{\rm th}$ percentiles of the distribution in bins of $D$ chosen to contain a similar number of galaxies. The horizontal error bars of the large grey open squares indicate the range of distances included in each bin. The vertical error bars of the large grey open squares indicate the RMS of the median.} \label{vs_dist}
\end{figure}

The parameters \lambdae\ and \lambdaee\ are normalized in the following way for visualization purposes:

\begin{equation}\label{eqn_lambdaec}
\lambda^*_{Re}=\frac{\lambda_{Re}}{\sqrt{\epsilon_{e}}}
\end{equation}

\begin{equation}\label{eqn_lambdaeec}
\lambda^*_{Re/2}=\frac{\lambda_{Re/2}}{\sqrt{\epsilon_{e/2}}}
\end{equation}

Figure \ref{vs_dist} shows \lambdaec\ and \lambdaeec\ as a function of projected clustercentric distance. The segregation between SR and FR is done following Equation \ref{eqn_lam_e}. Eight out of the 11 SR dEs are located in the inner $2^{\circ}$ of the Virgo cluster ($D \leq 0.4$~\Rvirial), two SR dEs are located at $D\sim 3^{\circ}$ ($D\sim 0.6$~\Rvirial), and the remaining SR dE, VCC~33, is located at $D\sim 5.4^{\circ}$ ($D\sim$~\Rvirial). The FR dEs are found at all radii within the Virgo cluster. The weighted average \lambdaec~ for FR dEs with $D<0.4$~\Rvirial is $0.41\pm0.03$, while the weighted average  \lambdaec~ for FR dEs with $D>$~\Rvirial~ is $0.67\pm0.03$. 
Spearman and Pearson statistical tests indicate with $99.92\%$ and $99.97\%$ confidence levels that there is a correlation between \lambdaec~ and $D$, respectively, and with $99.82\%$ and $99.77\%$ confidence levels that there is a correlation between \lambdaeec~ and $D$, respectively. This suggests that FR dEs in the outer parts of the cluster are rotating faster than FR dEs in the inner parts of the cluster.

The \ATLAS~ slow rotating ETGs are also preferentially located in the inner $2^{\circ}$ of the Virgo cluster ($D<0.4$~\Rvirial), although there are five exceptions at a distance of $4^{\circ} < D < 8^{\circ}$ (0.7~\Rvirial$<D<$1.5~\Rvirial). The Spearman and Pearson statistical tests do not find any correlation between \lambdaec, or \lambdaeec\ and the projected clustercentric distance for the \ATLAS~ galaxies (confidence levels of $97.58\%$, $91.74\%$, $90.35\%$, and $83.13\%$, respectively). However, note that the \ATLAS~ sample of Virgo cluster ETGs is distributed across all substructures of the Virgo cluster, while the SMAKCED dEs are nearly all in the Virgo cluster A (see Section \ref{sample}). Moreover, all of the \ATLAS~ ETGs have higher masses than the SMAKCED dEs, which could explain their lower sensitivity to the hot gas density \citep[mainly measured by the distance from M87;][]{Schindler99}, but higher sensitivity to the density of surrounding galaxies \citep[see the correlation between the kinematics and the local galaxy density for \ATLAS~ ETGs found by][]{Cappellari11}. In addition, \citet{Boselli14} find that the properties of Virgo cluster dEs do not correlate with the local galaxy density.

Dwarf early-type galaxies with and without underlying disky structures are present at all projected distances from the center of the Virgo cluster. However, applying binomial statistics, the fraction of slow rotating dEs with underlying disky structures is $0.27\pm0.13$ (3 out of 11), while the fraction of fast rotating dEs with underlying disky structures is $0.61\pm0.09$ (17 out of 28). These fractions suggest that underlying disky structures are more commonly found in fast rotating dEs than in slow rotating dEs.


\section{Comparison of the structure and stellar kinematics of dEs and low mass star-forming galaxies}\label{Sm_comp}

\begin{figure}
\centering
\includegraphics[angle=-90,width=8.5cm]{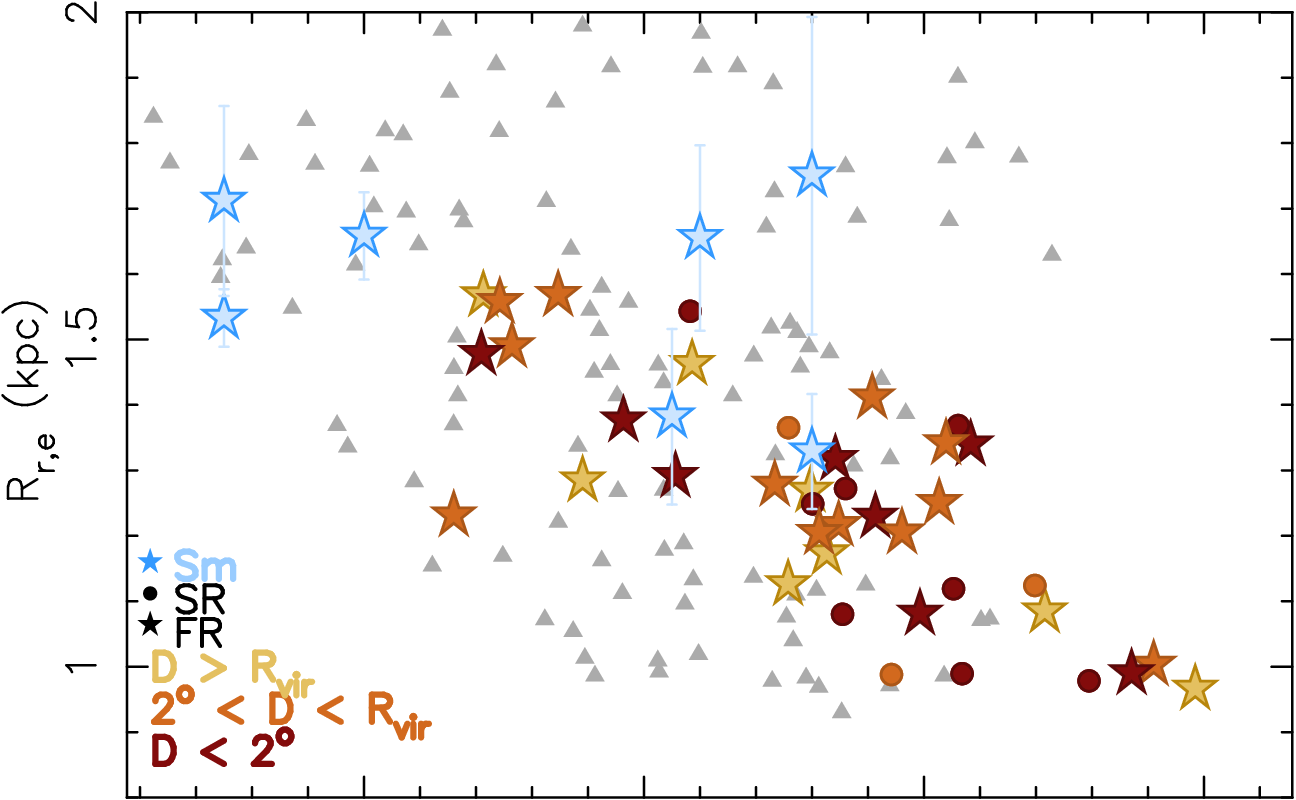}
\includegraphics[angle=-90,width=8.5cm]{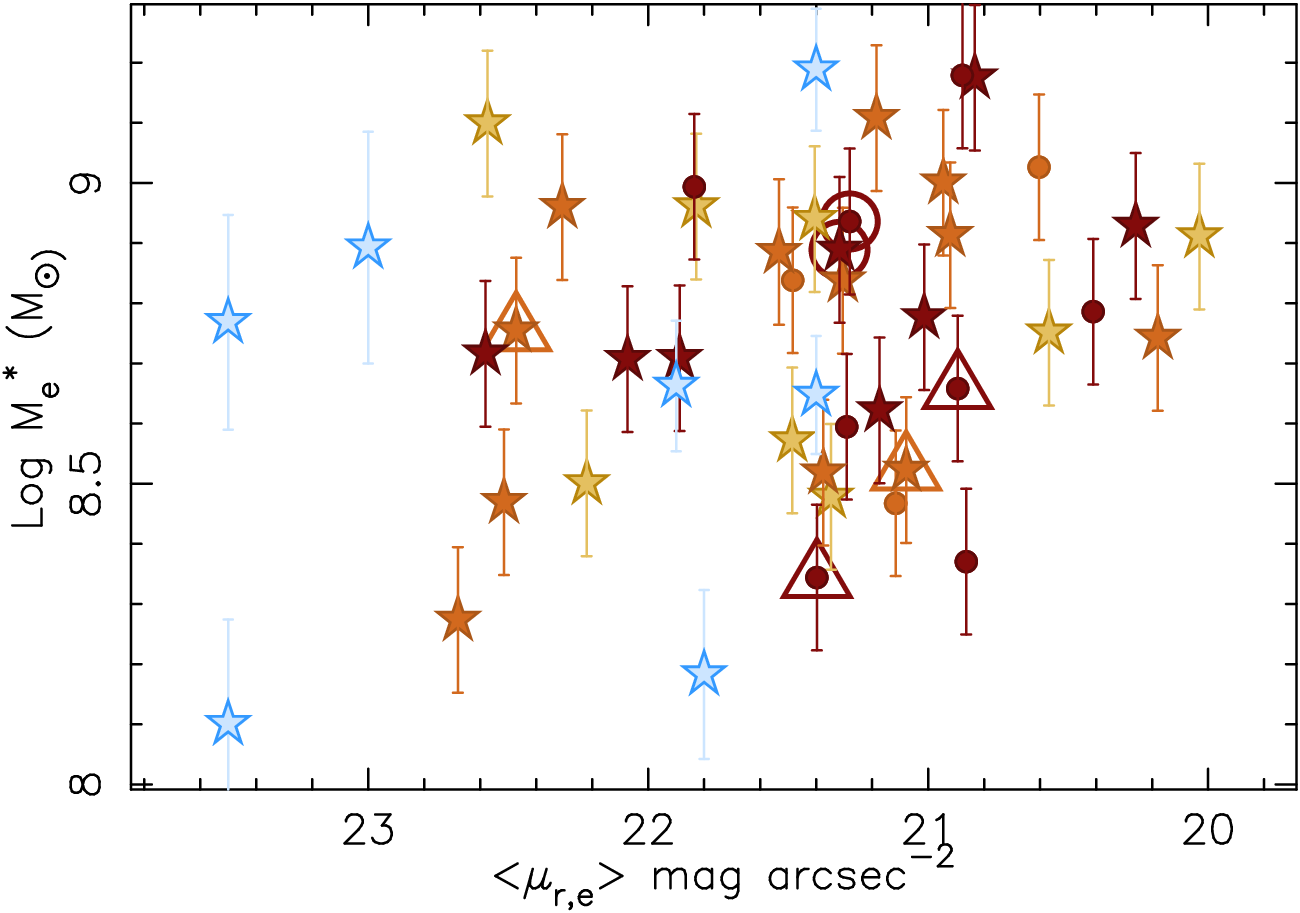}
\caption{{\bf Upper panel:} Half-light radius as a function of surface brightness, both in the $r$ band. Symbols' shapes are as in Figure \ref{vs_dist}. Yellow, orange, and red symbols are the SMAKCED dEs at different distances from the center of the Virgo cluster. Blue asterisks are the Sm galaxies from \citet{Adams14}. Grey triangles are Virgo cluster dEs in the magnitude range $-19.0 < M_r < -16.0$ \citep{Lisk07,Janz08}. {\bf Lower panel:} Stellar mass within the \Reff~ as a function of surface brightness in the $r$ band. The stellar masses for the SMAKCED dEs are calculated as described in Section \ref{kin_measurements} and for the Sm galaxies as described by \citet{Adams14}. The initial mass function considered for the SMAKCED dEs is Kroupa \citep{KroupaIMF} while for the Sm galaxies is Chabrier \citep{ChabrierIMF}. The Sm galaxies presented here tend to have slightly larger optical sizes than the SMAKCED dEs but are in the same range of stellar masses.}
\label{SB-size}
\end{figure}

We compare the structural properties and stellar kinematics of the SMAKCED dEs with the seven dwarf star-forming galaxies from \citet{Adams14}. None of these samples are complete. The SMAKCED dEs were targeted to preferentially have high surface brightness (Paper~II), while the Sm galaxies were targeted to preferentially have low surface brightness \citep{Adams14}. Figure \ref{SB-size} shows the size-surface brightness and stellar mass-surface brightness relations for both samples. While the Sm galaxies have slightly larger optical sizes than the dEs, both samples are in the same range of stellar masses.

\begin{figure}
\centering
\includegraphics[angle=-90,width=8.5cm]{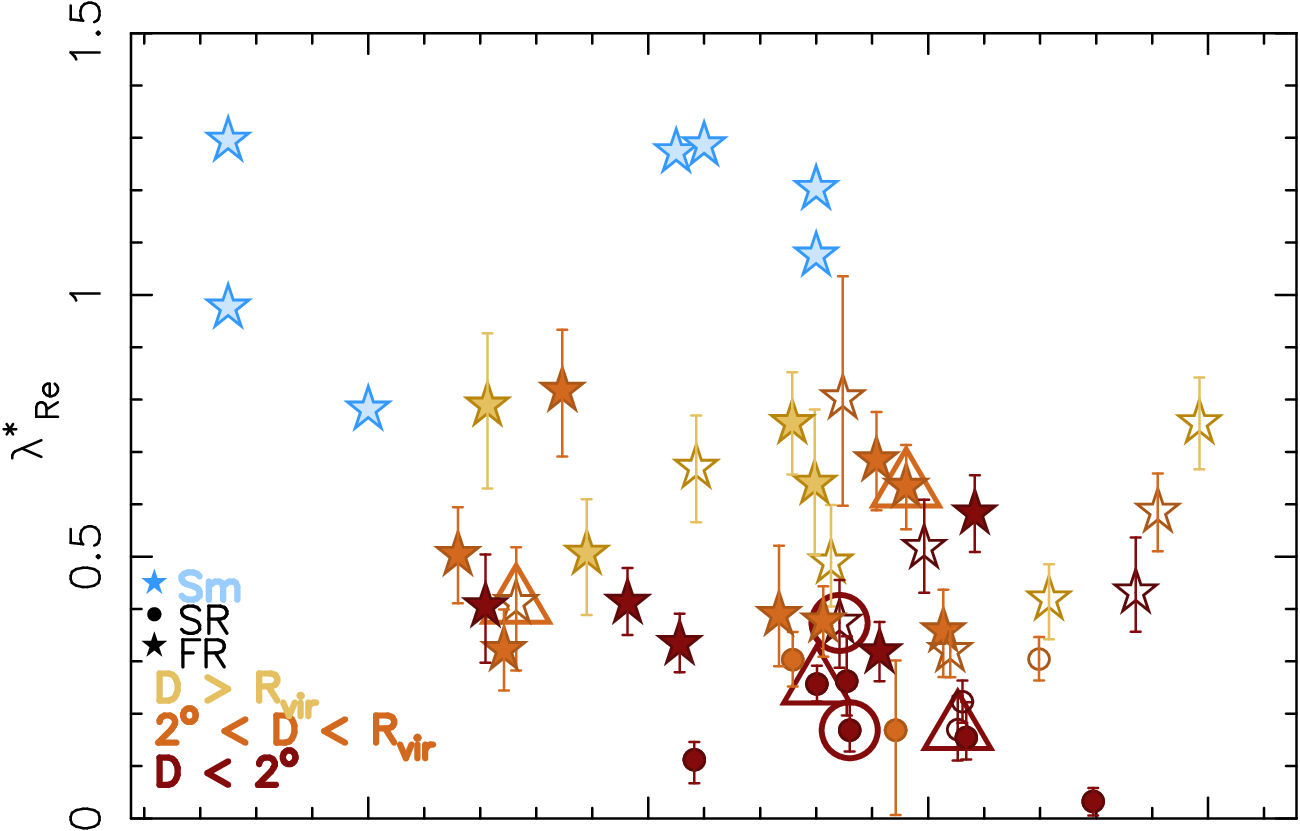}
\includegraphics[angle=-90,width=8.5cm]{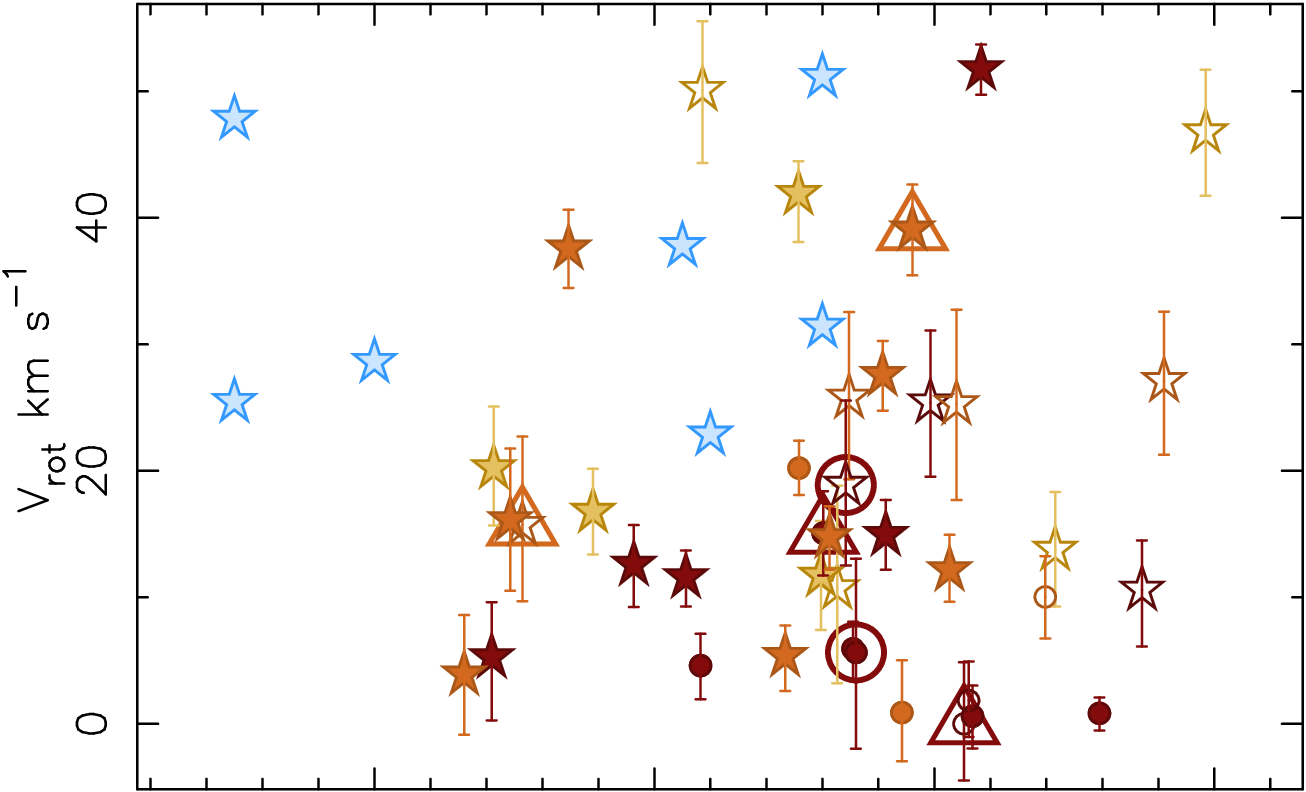}
\includegraphics[angle=-90,width=8.5cm]{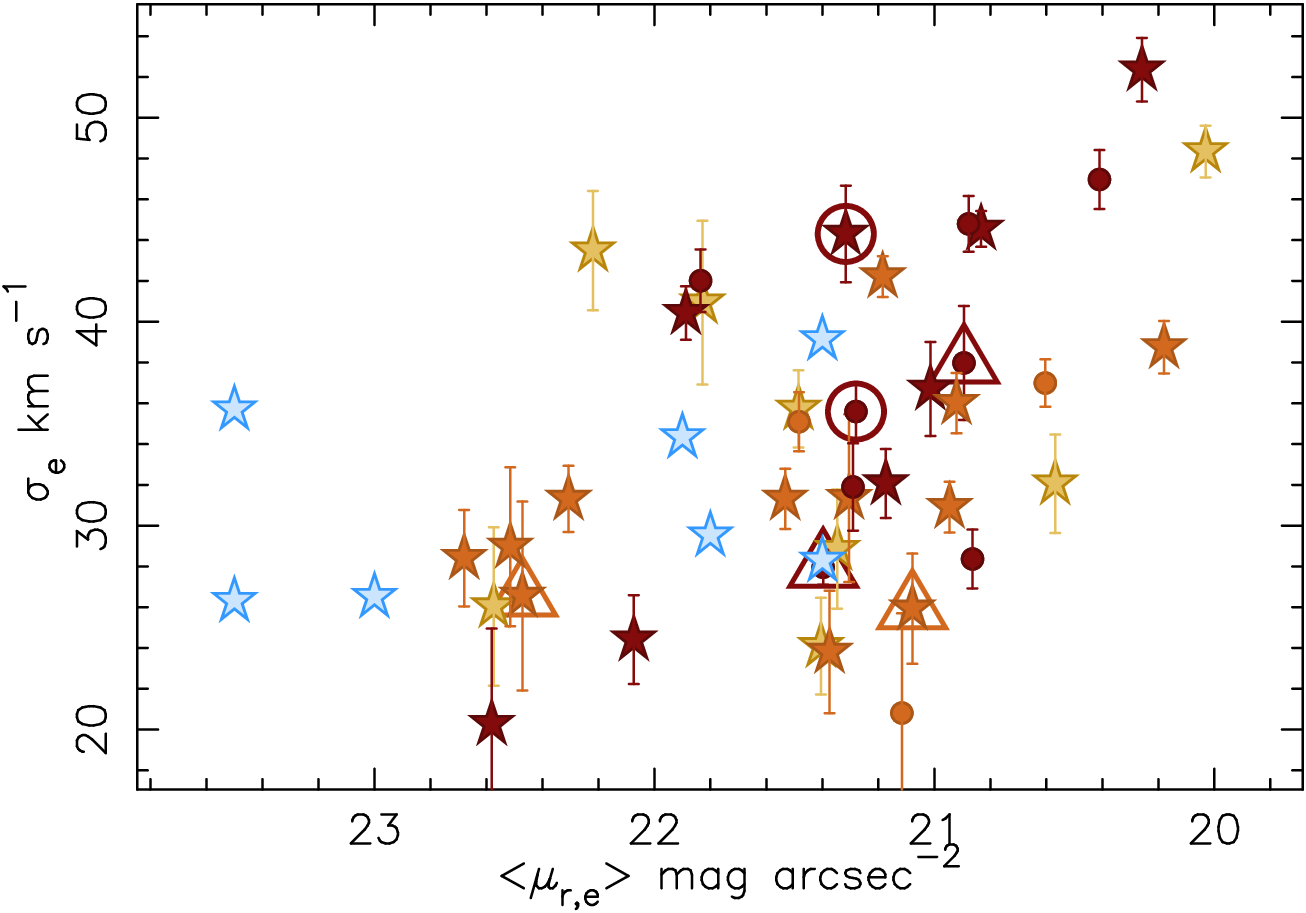}
\caption{{\bf Upper panel:} Stellar angular momentum \lambdaec\ as a function of surface brightness in the $r$ band. Symbols and colors are as in Figure \ref{SB-size}. {\bf Middle panel:} Stellar rotation speed \Vrot\ as a function of surface brightness in the $r$ band. The stellar rotation speed \Vrot\ for the Sm galaxies is calculated as described in Section \ref{kin_measurements} for the SMAKCED dEs. {\bf Lower panel:} Stellar velocity dispersion \sigmae\ as a function of surface brightness in the $r$ band. While the Sm galaxies tend to have higher \lambdaec\ and \Vrot\ than the dEs, they both have similar stellar velocity dispersions.}
\label{dE_Sm}
\end{figure}

\begin{figure}
\centering
\includegraphics[angle=-90,width=8.5cm]{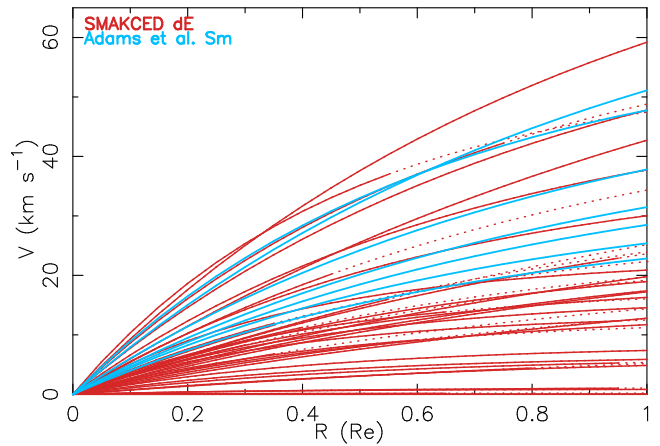}
\caption{Best fit {\it Polyex} function for the long-slit stellar rotation curves of the SMAKCED dEs and the long-slit stellar rotation curves of low luminosity Sm galaxies from \citet{Adams14}. Dotted lines indicate the regions where the kinematic information is not available.}
\label{polyex}
\end{figure}

The stellar kinematics of the Sm galaxies are derived from the integral field spectroscopy presented in \citet{Adams14}. To compare the \Vrot\ and $\sigma_e$ with the values obtained for the SMAKCED dEs, we extract the one dimensional rotation curve in a simulated long-slit along the major axis of the Sm galaxies. We fit the {\it Polyex} analytic function in the same way we did for the SMAKCED dEs (see Paper~II), and calculate the \Vrot\ as the velocity of the best fit {\it Polyex} function at the \Reff. The stellar velocity dispersion $\sigma_e$ for the Sm galaxies is the uncertainty-weighted average of $\sqrt{V^2+\sigma^2}$ calculated in each spatial resolution element with $R\leqslant$\Reff\ along the simulated long-slit described above.
Figure \ref{dE_Sm} shows the stellar kinematics as a function of surface brightness, and Figure \ref{polyex} shows the best fit {\it Polyex} function for the long-slit stellar rotation curves of the SMAKCED dEs and the Sm galaxies of \citet{Adams14}.

The dEs and Sm galaxies follow the same structural trends \citep[as previously found by ][]{Boselli08b}. The stellar rotation of the Sm galaxies of \citet{Adams14} overlap with the most rapid rotating SMAKCED dEs. In addition, these Sm galaxies overlap with the SMAKCED dEs that have the lowest stellar velocity dispersion. Thus, the normalized specific stellar rotation \lambdaec~ is higher for these Sm galaxies than for the SMAKCED dEs. There are also some structural and stellar kinematic similarities between blue compact dwarf galaxies and dEs \citep{Koleva14}. 


\section{Discussion}\label{discussion}

Due to the low masses ($M_e^* \sim 10^9$~M$_{\odot}$) of dEs they are expected to be strongly affected by the environment. If that is the case, some of the internal properties of the dEs should correlate with fundamental properties of the cluster. In this work, we have studied the kinematics of 39 dEs as a function of projected distance from the center of the Virgo cluster.

\subsection{Correlations with projected clustercentric distance}

In \citet{etj09} we investigated the dynamical support and luminosity-weighted ages of a subsample of the SMAKCED dEs as a function of projected distance from the center of the Virgo cluster. We found a hint of correlation between the angular momentum, measured by \vs, and the projected clustercentric distance. In addition, we found that the luminosity-weighted stellar ages of the inner 4$''$ of the dEs are preferentially younger in the outer parts of the Virgo cluster.

In this paper, we improve the analysis of \citet{etj09} in three ways: (1) we more than double the sample of Virgo cluster dEs to increase the statistics; (2) instead of using the highest rotation speed measured for each dE, we fit the {\it Polyex} function to the rotation curve, and measure the rotation speed at the \Reff~ for all dEs, using the full rotation curve; and (3) we complement the analysis of \vs~ with the \lambdae\ and \lambdaee\ parameters, which measure the specific stellar angular momentum within an aperture (see Equation \ref{lamb_eqn}).

To study the radial gradients of galaxy properties within the cluster we need the three-dimensional (3D) distance of each galaxy to the center of the cluster. Unfortunately, the 3D distance of any dE to the center of the cluster cannot be easily derived because of two main reasons: (1) the line-of-sight depth of the Virgo cluster is $\sim 4$~Mpc including all the substructures. The majority of the SMAKCED dEs are part of the Virgo A cluster, centered in M87, which has a line-of-sight depth of 1.5~Mpc \citep{Mei07}. The uncertainties in the line-of-sight distances of the brightest dEs measured with surface brightness fluctuations, the most accurate technique to obtain distances, are from 0.5 to over 1~Mpc, while the distance for the faintest dEs cannot be measured \citep{Mei07}; (2) in a 3D structure, the orbital parameters of the dEs are more important than their current position. Because dEs move in elliptical orbits, they do not spend the same amount of time in each part of their orbit. Due to the difficulty to overcome these two effects, we are limited to projected distances to look for environmental effects.

The projected  clustercentric distance can smear out any correlation present, i.e. dEs could appear close to the center of the cluster by projection effects making the correlation weaker. In addition, there can be back-splash galaxies also contributing to make the correlation weaker, i.e. galaxies that were near the center of the cluster and are currently in a part of their orbit that is further away. 

The kinematic properties of the SMAKCED dEs are related to the presence or absence of subtle disky structures found in high-pass filtered optical images by \citet{Lisk06a,Lisk07}. The dEs with high stellar angular momentum show these subtle underlying disky structures while dEs with low stellar angular momentum preferentially lack these structures. However, we do not find any evidence for a correlation between the number of components in which the $H$ band surface brightness profiles of the dEs are decomposed by \citep{Janz14} and their stellar angular momentum.

The stellar angular momentum increases with the projected clustercentric distance \citep[confirming our previous result in][]{etj09}. However, \citet{Rys13} did not find any correlation between \lambdae~  and the distance to the center of the cluster. Since our \lambdae~ values agree with those of Ry\'s et al., this discrepancy is probably due to the larger sample used here and their attempt to use 3D distances. 

The fast rotators in the outer parts of the Virgo cluster are rotating faster than the fast rotators in the inner parts of the cluster. We find eleven slow rotators in our sample. Eight of them are located at a projected clustercentric distance of $D<2^{\circ}$ ($D<0.4$~\Rvirial), two of them are located at $D\sim 3^{\circ}$ ($D\sim 0.6$~\Rvirial), while the remaining slow rotator is found at a projected clustercentric distance of $D\sim 5.4^{\circ}$ ($D\sim$~\Rvirial) and does not have any massive close neighbor within a radius of 150~kpc. Eight out of the 11 slow rotators do not have any underlying disky structures. However, two of them, even though they are located in projection very close to M87, have emission lines. One of the two central SR dEs with emission lines and underlying disky structures is VCC~1684, a very flattened galaxy and a good candidate to be a $2\sigma$ galaxy \citep[i.e. a galaxy with two peaks in the velocity dispersion profile likely due to the presence of two counter rotating disks; see Section \ref{vs_section} and][]{Kraj11}.

\subsection{Possible Formation Scenarios}

Our findings are consistent with a scenario in which late-type star-forming galaxies fall into the Virgo cluster and are transformed into the dEs that we observe today. The dynamical, structural, and stellar population properties analyzed in this series of papers provide new constraints on the physical mechanisms that can cause such transformation and, thus, on which galaxies are the progenitors of dEs. 
The two most discussed physical mechanisms to explain the transformation of late-type star-forming galaxies into dwarf early-type (quenched) galaxies are ram pressure stripping and galaxy harassment. 

Ram pressure stripping, hydrodynamical interaction with the hot intracluster medium, removes the gas of the galaxy and quenches the star formation on a very short time scale. The gas removal takes $100-200$~Myr and the galaxy looks red and quiescent in $\sim 1$~Gyr \citep[][]{Boselli08a}. The resulting galaxy has a slightly lower surface brightness in the optical bands (by $\sim1$~mag) because of the quenching of the star formation, but the \Reff\ and the stellar mass remain basically the same \citep{Boselli08b}. As ram pressure stripping does not directly affect the stars, the galaxy conserves its stellar rotation. After a long time in the cluster, the galaxy goes through the center of the cluster several times. These multiple fast passages heat up the galaxy and make it lose its rotation \citep{Boselli08a,Boselli08b}. Ram pressure stripping is effective at all distances from the center of the cluster, but it is more efficient in regions with denser intergalactic medium.

Harassment, multiple gravitational interactions of the dE progenitor with other cluster members and with the potential of the cluster, affects both the gas and the stars \citep[e.g.][]{Moore98,Mast05}. This mechanism is efficient only when the galaxy goes through very populated and dense regions many times. The typical crossing time of the Virgo cluster is $\sim 1.7$~Gyr \citep[e.g.][]{BG06}, thus it takes several Gyrs to make strong changes in the structure and kinematics of the progenitor galaxy. 
Large amounts of gas and stars are removed in the gravitational interactions. Therefore, the remnant galaxy is smaller and less massive than its progenitor \citep{Moore98,Mast05,Boselli08b}. Harassment also decreases the strength of rotation, generating even non-rotating galaxies under very extreme conditions, and changes the internal structure of the progenitor into a less disky or even a pure spheroidal galaxy \citep{Mast05}. This mechanism is most effective where the density of neighbors is high, i.e. in the central parts of the cluster. In the outer regions of Virgo, the density of neighbors is so low that the progenitors are not expected to be significantly affected by harassment \citep{Smith10}. 

The observed trend between the specific stellar angular momentum of dEs and the clustercentric distance suggests that the environment is removing the rotation in these galaxies. Given that the faint end of the red sequence seems to have been formed in the last few Gyrs \citep{Stott07,deLucia07}, that quenched dwarf galaxies are not found in isolation \citep{Gavazzi10,Geha12}, and that the infall rate of late-type galaxies in the Virgo cluster is large \citep[$300-400$ late-type galaxies per Gyr][]{Boselli08a,Gavazzi13}, the environmental mechanism transforming these infalling late-types into quenched galaxies must be fast. While ram pressure stripping is effective in a very short time scale, i.e. less than 1~Gyr, harassment takes several Gyrs to fully quench and transform the galaxies. This suggests that ram pressure stripping might be the main mechanism transforming freshly infalling low luminosity late-type galaxies into dEs. However, the galaxies that entered the cluster in its early epochs may have been affected by other processes, too.

 In this scenario, a low luminosity star-forming galaxy enters the Virgo cluster and its gas is rapidly removed by ram pressure, in $100-200$~Myr, but its internal kinematics and structure are preserved appearing like the FR dEs with subtle disky structures seen in the outer parts of the cluster. After being in the cluster for a long time, i.e. a few Gyrs, and passing through its center several times, the galaxy is heated up, loses stellar rotation, and the internal disky structures become less prominent, i.e. the spiral arms begin to disappear. In these several passes through the center of the cluster and in addition to ram pressure, gravitational interactions with other cluster members and with the cluster potential well, i.e. harassment and tidal stirring, can also play a role in the transformation of the internal properties of the galaxy  (Benson et al. 2014, Boselli $\&$ Gavazzi 2015). This may explain the properties of the FR dEs with subtle disky structures seen in the inner parts of the cluster which will eventually be slow rotating dEs. These fast rotators found in the inner parts of the cluster rotate more slowly than the fast rotators found in the outer parts of the cluster.
Cosmological simulations predict that clusters were formed by the accretion of small groups of galaxies \citep[e.g.][]{Springel05}. The small velocity dispersion of these groups favored gravitational interactions between the group members in the early epochs of the cluster formation. This scenario explains the presence of slow and non-rotating dEs in the center of the cluster as well as the presence of kinematically decoupled cores in some of these dEs (see Paper~I).

This transformation scenario is supported by the fact that slow rotating dEs are gravitationally bound to the cluster while fast rotating dEs are mainly unbound systems and have a radial velocity distribution within the cluster that is very similar to that of late-type galaxies \citep{Boselli14}. This suggests that slow rotating dEs, which do not have underlying disky structures, may have been in the cluster for much longer than fast rotating dEs, which do have underlying disky structures, and may have entered the cluster more recently.

Four of the SMAKCED dEs have some residual star formation in their central regions. As discussed in Paper~I for the formation of kinematically decoupled cores in dEs, the accretion of gas from another galaxy is unlikely inside clusters due to the rapid relative velocities of galaxies \citep[e.g.][]{BinnTrem87,BG06}. Thus, this gas might come from its star-forming progenitor which was not fully stripped by ram pressure. Simulations expect this to happen in the core of some galaxies where the potential well is deeper \citep{Boselli08a}.

To observationally probe this formation scenario it is important to compare the stellar kinematics of dEs with those of star-forming galaxies. However, the study of the stellar kinematics of star-forming galaxies is still on its infancy. The first papers devoted to such a study are \citet{Koleva14} and \citet{Adams14}. The analysis done by \citet{Koleva14} and our comparison with \citet{Adams14} in Section \ref{Sm_comp} show some structural and kinematic similarities between star-forming and quenched dwarf galaxies.

\section{Summary and Conclusions}\label{concl}

In this work, we analyze the stellar kinematics of dEs in the Virgo cluster as a function of projected clustercentric distance, which can serve as a proxy for the impact that the environment may have had on galaxies. We use intermediate resolution optical spectroscopy of 39 dEs in the Virgo cluster, which is the largest spectroscopic sample of Virgo cluster dEs observed and analyzed homogeneously up to date.

We find 11 slow rotators (SR) and 28 fast rotators (FR) within the Virgo cluster dEs. In addition, the range of values of \lambdae\ found for the dEs (\lambdae$\leq 0.4$) is smaller than the range of \lambdae\ values found for fast rotating ETGs \citep[\lambdae$\leq 0.8$; as hinted by][]{Rys14}. A quantitative comparison between these ETGs and dEs is complex because of the nature of the samples \citep[see][]{Janz14}. While the ATLAS$^{3D}$ sample of ETGs is volume complete, the SMAKCED sample of dEs is representative of dEs in the magnitude range $-19.0 < M_r < -16.0$ but it is not complete; these two samples of galaxies do not overlap in mass; and while the ATLAS$^{3D}$ sample consists mainly of group and field ETGs, the SMAKCED sample consists only of cluster dEs. However, finding that slow rotators dominate within the most massive ETGs and are also found within the dE galaxy class suggests that the process forming slow rotators might not be the same for galaxies of different masses.

Dwarf early-type galaxies with strong rotation, quantified by \vse, \lambdae, and \lambdaee, contain subtle disky structures, such as spiral arms, disks, and irregular features, seen in high-pass filtered optical images, while dEs with low rotation do not exhibit such subtle disky structures. We also find a correlation between the stellar rotation and the projected clustercentric distance. The dEs in the outer parts of the cluster rotate faster than the dEs in the inner parts. The SR dEs are found mainly around M87, the galaxy at the center of the Virgo cluster.

The properties of dEs in the Virgo cluster can be explained by a scenario in which  late-type star-forming galaxies are transformed by the environment. In this scenario, the star-forming galaxy falls into the cluster, its gas is removed very rapidly, in $100-200$~Myr, by ram pressure stripping, its star formation is quenched and in $\sim 1$~Gyr of passive evolution its stellar population looks red. The resulting galaxy is likely to have the properties of the FR dEs with some subtle disky structures that we find in the outer parts of the Virgo cluster. As the galaxy passes several times through the center of the cluster, the galaxy is heated up and begins to lose its rotation and subtle structures eventually becoming a pure spheroid with no rotation as the ones seen in the center of Virgo.

Those galaxies that entered the cluster in the early stages of its formation may have been affected by other mechanisms in addition to ram pressure stripping. As the cluster was mainly formed by the accretion of small groups of galaxies \citep{Springel05} in its early stages, gravitational interactions between group members may have been frequent. This scenario is supported by the kinematically decoupled cores observed in some of the SMAKCED dEs (see Paper~I ).

The structural properties and stellar kinematics of the Virgo cluster dEs share some similarities with star-forming dwarf galaxies (see Section \ref{Sm_comp}). However, stellar kinematics of complete samples of dEs and star-forming galaxies are needed to conclude whether they share a common origin.

\acknowledgments

The authors thank the referee for useful suggestions that improved this manuscript.
E.T. acknowledges the financial support of the Fulbright Program jointly with the Spanish Ministry of Education. PG acknowledges the NSF grant AST-1010039. T.L. was supported within the framework of the Excellence Initiative by the German Research Foundation (DFG) through the Heidelberg Graduate School of Fundamental Physics (grant number GSC 129/1). RFP, GvdV, JFB, EL, and HS acknowledge the DAGAL network from the People Programme (Marie Curie Actions) of the European Union’s Seventh Framework Programme FP7/2007-2013/ under REA grant agreement number PITN-GA-2011-289313. G.H. acknowledges support by the FWF project P21097-N16. J.J. thanks the ARC for financial support (DP130100388). This work has made use of the GOLDMine database \citep{GOLDMine} and the NASA/IPAC Extragalactic Database (NED.\footnote{http://ned.ipac.caltech.edu})

\bibliographystyle{aa}
\bibliography{references}{}


\end{document}

%% file: kinematics_table.tex
\begin{table*}
\begin{center}
\caption{Kinematic and Structural Properties of the SMAKCED dEs. \label{tabledEs}}
{\renewcommand{\arraystretch}{1.}
\resizebox{18cm}{!} {
\begin{tabular}{|c|c|c|c|c|c|c|c|c|c|c|c|}
\hline \hline
Galaxy              &  RA  &  Dec  & $D$ & $R_{\rm max}/$\Reff & \vs & \lambdaeed & \lambdaed & $\epsilon_{\rm e/2}$ & $\epsilon_e$ & {\bf FR/SR (\Reff$/2$)} &{\bf FR/SR (\Reff)} \\
                        & hh:mm:ss& dd:mm:ss &   deg  &    &    &    &    &    &   &  & \\
(1)                    & (2)            &  (3)           &  (4)     &(5)&(6)&(7)&(8)&(9) &(10) & (11) & (12) \\ 
\hline
VCC0009 & 12:09:22.25 & +13:59:32.74& 5.56  &  1.0  & 0.74$^{+0.15}_{-0.15 }$ & 0.32$^{+0.06}_{-0.06 }$ & 0.54$^{+0.07}_{-0.08 }$ & 0.212$\pm$0.003 & 0.189$\pm$0.002 &FR&FR\\
VCC0021 & 12:10:23.15 & +10:11:19.04& 5.53  &  0.4 & (0.37$^{+0.19}_{-0.21}$) & 0.36$^{+0.08}_{-0.05}$ & (0.46$^{+0.12}_{-0.10}$) & 0.360$\pm$0.006 & 0.357$\pm$0.005 &FR&FR\\
VCC0033 & 12:11:07.79 & +14:16:29.19& 5.24  &  1.0  & 0.01$^{+0.14}_{-0.15 }$ & 0.09$^{+0.10}_{-0.05 }$ & 0.11$^{+0.08}_{-0.10 }$ & 0.211$\pm$0.006 & 0.168$\pm$0.004 &SR&SR \\
VCC0170 & 12:15:56.34 & +14:26:00.33& 4.17  &  0.7 & (0.39$^{+0.19}_{-0.21}$) & 0.30$^{+0.09}_{-0.10}$ & (0.39$^{+0.11}_{-0.13}$) & 0.392$\pm$0.003 & 0.368$\pm$0.002 &FR&FR \\
VCC0308 & 12:18:50.90 & +07:51:43.38& 5.42  &  1.3  & 0.46$^{+0.10}_{-0.11 }$ & 0.17$^{+0.03}_{-0.05 }$ & 0.31$^{+0.06}_{-0.05 }$ & 0.102$\pm$0.003 & 0.097$\pm$0.003 &FR&FR\\
VCC0389 & 12:20:03.29 & +14:57:41.70& 3.71  &  1.0  & 0.41$^{+0.05}_{-0.05 }$ & 0.19$^{+0.05}_{-0.03 }$ & 0.25$^{+0.05}_{-0.05 }$ & 0.182$\pm$0.002 & 0.198$\pm$0.001 &FR&FR\\
VCC0397 & 12:20:12.18 & +06:37:23.51& 6.34  &  1.2  & 1.18$^{+0.07}_{-0.08 }$ & 0.41$^{+0.04}_{-0.03 }$ & 0.62$^{+0.02}_{-0.02 }$ & 0.234$\pm$0.002 & 0.277$\pm$0.002 &FR&FR\\
VCC0437 & 12:20:48.10 & +17:29:16.00& 5.67  &  0.8 & (1.23$^{+0.12}_{-0.13}$) & 0.49$^{+0.05}_{-0.06}$ & (0.60$^{+0.12}_{-0.12}$) & 0.314$\pm$0.002 & 0.324$\pm$0.002 &FR&FR\\
VCC0523 & 12:22:04.14 & +12:47:14.60& 2.21  &  2.0  & 0.65$^{+0.02}_{-0.02 }$ & 0.35$^{+0.03}_{-0.03 }$ & 0.54$^{+0.03}_{-0.03 }$ & 0.252$\pm$0.002 & 0.256$\pm$0.002 &FR&FR\\
VCC0543 & 12:22:19.54 & +14:45:38.59& 3.17  &  1.5  & 0.58$^{+0.03}_{-0.03 }$ & 0.27$^{+0.02}_{-0.04 }$ & 0.31$^{+0.04}_{-0.04 }$ & 0.428$\pm$0.001 & 0.436$\pm$0.001 &SR&SR\\
VCC0634 & 12:23:20.01 & +15:49:13.25& 3.90  &  1.6  & 1.39$^{+0.07}_{-0.07 }$ & 0.51$^{+0.02}_{-0.02 }$ & 0.52$^{+0.04}_{-0.05 }$ & 0.151$\pm$0.002 & 0.164$\pm$0.002 &FR&FR\\
VCC0750 & 12:24:49.58 & +06:45:34.49& 5.82  &  1.5  & 0.38$^{+0.06}_{-0.05 }$ & 0.13$^{+0.07}_{-0.06 }$ & 0.28$^{+0.05}_{-0.05 }$ & 0.108$\pm$0.003 & 0.125$\pm$0.003 &SR&FR\\
VCC0751 & 12:24:48.30 & +18:11:47.00& 5.99  &  0.4 & (0.39$^{+0.08}_{-0.08}$) & 0.32$^{+0.05}_{-0.05}$ & (0.41$^{+0.08}_{-0.09}$) & 0.388$\pm$0.001 & 0.394$\pm$0.001 &FR&FR\\
VCC0781 & 12:25:15.17 & +12:42:52.59& 1.42  &  0.7 & (0.01$^{+0.10}_{-0.10}$) & 0.07$^{+0.05}_{-0.04}$ & (0.14$^{+0.06}_{-0.05}$) & 0.238$\pm$0.004 & 0.282$\pm$0.003 &SR&FR\\
VCC0794 & 12:25:22.10 & +16:25:47.00& 4.26  &  1.0  & 0.51$^{+0.16}_{-0.16 }$ & 0.28$^{+0.09}_{-0.06 }$ & 0.34$^{+0.07}_{-0.07 }$ & 0.359$\pm$0.002 & 0.446$\pm$0.001 &FR&FR\\
VCC0856 & 12:25:57.93 & +10:03:13.54& 2.63  &  0.4 & (0.82$^{+0.23}_{-0.24}$) & 0.25$^{+0.08}_{-0.07}$ & (0.34$^{+0.11}_{-0.10}$) & 0.055$\pm$0.004 & 0.072$\pm$0.003 &FR&FR\\
VCC0917 & 12:26:32.39 & +13:34:43.54& 1.59  &  1.7  & 0.02$^{+0.04}_{-0.06 }$ & 0.10$^{+0.03}_{-0.03 }$ & 0.13$^{+0.05}_{-0.03 }$ & 0.270$\pm$0.002 & 0.285$\pm$0.002 &SR&SR\\
VCC0940 & 12:26:47.07 & +12:27:14.17& 1.00  &  1.1  & 0.29$^{+0.02}_{-0.04 }$ & 0.15$^{+0.02}_{-0.02 }$ & 0.18$^{+0.02}_{-0.02 }$ & 0.075$\pm$0.002 & 0.120$\pm$0.002 &FR&FR\\
VCC0990 & 12:27:16.94 & +16:01:27.92& 3.74  &  0.5 & (0.70$^{+0.06}_{-0.09}$) & 0.35$^{+0.03}_{-0.03}$ & (0.44$^{+0.08}_{-0.08}$) & 0.200$\pm$0.002 & 0.232$\pm$0.002 &FR&FR\\
VCC1010 & 12:27:27.39 & +12:17:25.09& 0.84  &  2.3  & 1.16$^{+0.03}_{-0.03 }$ & 0.46$^{+0.01}_{-0.02 }$ & 0.60$^{+0.01}_{-0.01 }$ & 0.368$\pm$0.002 & 0.429$\pm$0.002 &FR&FR\\
VCC1087 & 12:28:14.90 & +11:47:23.58& 0.88  &  1.9  & 0.11$^{+0.04}_{-0.03 }$ & 0.06$^{+0.02}_{-0.02 }$ & 0.11$^{+0.03}_{-0.04 }$ & 0.423$\pm$0.002 & 0.380$\pm$0.001 &SR&SR\\
VCC1122 & 12:28:41.71 & +12:54:57.08& 0.74  &  1.3  & 0.47$^{+0.05}_{-0.05 }$ & 0.20$^{+0.04}_{-0.03 }$ & 0.32$^{+0.04}_{-0.04 }$ & 0.303$\pm$0.001 & 0.410$\pm$0.001 &SR&FR\\
VCC1183 & 12:29:22.51 & +11:26:01.73& 1.02  &  0.3 & (0.43$^{+0.13}_{-0.13}$) & 0.14$^{+0.04}_{-0.04}$ & (0.21$^{+0.05}_{-0.06}$) & 0.102$\pm$0.002 & 0.132$\pm$0.002 &FR&FR\\
VCC1261 & 12:30:10.32 & +10:46:46.51& 1.62  &  0.6 & (0.04$^{+0.07}_{-0.07}$) & 0.11$^{+0.03}_{-0.03}$ & (0.18$^{+0.04}_{-0.04}$) & 0.193$\pm$0.001 & 0.254$\pm$0.001 &SR&FR\\
VCC1304 & 12:30:39.90 & +15:07:46.68& 2.74  &  2.2  & 1.53$^{+0.16}_{-0.16 }$ & 0.53$^{+0.04}_{-0.03 }$ & 0.72$^{+0.02}_{-0.02 }$ & 0.429$\pm$0.003 & 0.525$\pm$0.002 &FR&FR\\
VCC1355 & 12:31:20.21 & +14:06:54.93& 1.73  &  1.0  & 0.30$^{+0.15}_{-0.20 }$ & 0.22$^{+0.07}_{-0.07 }$ & 0.30$^{+0.06}_{-0.07 }$ & 0.202$\pm$0.003 & 0.229$\pm$0.002 &FR&FR\\
VCC1407 & 12:32:02.73 & +11:53:24.46& 0.59  &  1.2  & 0.17$^{+0.05}_{-0.07 }$ & 0.12$^{+0.03}_{-0.04 }$ & 0.17$^{+0.05}_{-0.04 }$ & 0.158$\pm$0.003 & 0.164$\pm$0.002 &SR&SR\\
VCC1431 & 12:32:23.41 & +11:15:46.94& 1.19  &  0.6 & (0.20$^{+0.05}_{-0.07}$) & 0.13$^{+0.04}_{-0.03}$ & (0.21$^{+0.06}_{-0.04}$) & 0.108$\pm$0.001 & 0.093$\pm$0.001 &SR&FR\\
VCC1453 & 12:32:44.22 & +14:11:46.17& 1.87  &  1.7  & 0.16$^{+0.09}_{-0.10 }$ & 0.16$^{+0.03}_{-0.03 }$ & 0.12$^{+0.03}_{-0.02 }$ & 0.188$\pm$0.002 & 0.198$\pm$0.001 &SR&SR\\
VCC1528 & 12:33:51.61 & +13:19:21.03& 1.20  &  1.7  & 0.02$^{+0.03}_{-0.03 }$ & 0.01$^{+0.02}_{-0.02 }$ & 0.02$^{+0.02}_{-0.02 }$ & 0.213$\pm$0.001 & 0.206$\pm$0.001 &SR&SR\\
VCC1549 & 12:34:14.83 & +11:04:17.51& 1.57  &  0.4 & (0.69$^{+0.08}_{-0.08}$) & 0.24$^{+0.04}_{-0.04}$ & (0.32$^{+0.07}_{-0.07}$) & 0.147$\pm$0.002 & 0.155$\pm$0.002 &FR&FR\\
VCC1684 & 12:36:39.40 & +11:06:06.97& 1.94  &  1.1  & 0.54$^{+0.02}_{-0.02 }$ & 0.15$^{+0.01}_{-0.02 }$ & 0.32$^{+0.02}_{-0.01 }$ & 0.609$\pm$0.003 & 0.641$\pm$0.002 &SR&SR\\
VCC1695 & 12:36:54.85 & +12:31:11.93& 1.52  &  1.3  & 0.52$^{+0.07}_{-0.06 }$ & 0.23$^{+0.03}_{-0.02 }$ & 0.36$^{+0.04}_{-0.03 }$ & 0.316$\pm$0.002 & 0.307$\pm$0.002 &SR&FR\\
VCC1861 & 12:40:58.57 & +11:11:04.34& 2.79  &  1.3  & 0.14$^{+0.05}_{-0.04 }$ & 0.16$^{+0.03}_{-0.02 }$ & 0.12$^{+0.04}_{-0.03 }$ & 0.041$\pm$0.002 & 0.039$\pm$0.002 &FR&FR\\
VCC1895 & 12:41:51.97 & +09:24:10.28& 4.05  &  1.7  & 0.75$^{+0.13}_{-0.12 }$ & 0.29$^{+0.06}_{-0.06 }$ & 0.41$^{+0.05}_{-0.05 }$ & 0.481$\pm$0.004 & 0.492$\pm$0.003 &SR&FR\\
VCC1910 & 12:42:08.67 & +11:45:15.19& 2.88  &  0.6 & (0.27$^{+0.04}_{-0.03}$) & 0.12$^{+0.02}_{-0.02}$ & (0.19$^{+0.04}_{-0.03}$) & 0.150$\pm$0.003 & 0.161$\pm$0.002 &SR&FR\\
VCC1912 & 12:42:09.07 & +12:35:47.93& 2.82  &  0.6 & (0.70$^{+0.05}_{-0.05}$) & 0.22$^{+0.03}_{-0.03}$ & (0.30$^{+0.06}_{-0.06}$) & 0.280$\pm$0.002 & 0.367$\pm$0.002 &FR&FR\\
VCC1947 & 12:42:56.34 & +03:40:35.78& 9.22  &  0.6 & (0.97$^{+0.04}_{-0.03}$) & 0.43$^{+0.02}_{-0.02}$ & (0.53$^{+0.09}_{-0.09}$) & 0.186$\pm$0.002 & 0.203$\pm$0.001 &FR&FR\\
VCC2083 & 12:50:14.48 & +10:32:24.07& 5.16  &  0.8  & 0.14$^{+0.05}_{-0.05 }$ & 0.26$^{+0.05}_{-0.04 }$ & 0.31$^{+0.04}_{-0.04 }$ & 0.156$\pm$0.007 & 0.156$\pm$0.005 &FR&FR\\
\hline
\end{tabular}}
}
\end{center}
\tablecomments{Column 1: galaxy name. Columns 2 and 3: right ascension and declination in J2000. Column 4: projected distance between each dE and M87, considered to be the center of the Virgo cluster {\bf ($1^{\circ}\sim 0.29$~Mpc).} Column 5: maximum radial coverage of the rotation curve. Column 6: ratio between the rotation velocity measured at the \Reff~ and the velocity dispersion measured within the \Reff. We use Equation \ref{scale_vs_eqn} to transform \vs\ measured in long-slit data into the integrated \vse. Numbers within brackets indicate that the radial coverage of the rotation curve does not reach $\sim$~1\Reff, thus \Vrot\ is extrapolated up to the \Reff\ using the best fit {\it Polyex} fitting function to the rotation curve (see Paper~II for details). The parameters \Vrot\ and $\sigma_e$ can be found in Paper~II. Columns 7 and 8: specific stellar angular momentum measured  in the long-slit data within the \Reff$/2$ and the \Reff, respectively. We use Equation \ref{scale_lamb_eqn} to transform the long-slit measurements into the integrated \lambdaee\ and \lambdae\ values. Numbers within brackets indicate that the radial coverage of the rotation curve does not reach $\sim$~1\Reff, thus \lambdaed\ is estimated using the correlation between \lambdae\ and \lambdaee\ seen in Figure \ref{lambRevsRe2}. Columns 9 and 10: {\bf uncertainty-weighted mean ellipticities within the \Reff$/2$ and the \Reff, respectively. Columns 11 and 12: fast rotator (FR) or slow rotator (SR) classification based on the \lambdaee$-\epsilon_{\rm e/2}$, and \lambdae$-\epsilon_e$ relations from Equations \ref{eqn_lam_e} and \ref{eqn_lam_ee}.}}
\end{table*}